\documentclass[usegraphicx,usenatbib,useAMS]{emulateapj}
\usepackage{natbib}

\newcommand\Eqn[1]     {Eq.\,(\ref{#1})}
\newcommand\Eqns[2]    {Eqs\,(\ref{#1}) and~(\ref{#2})}

\newcommand\Fig[1]     {Fig.\,{\ref{#1}}}

\newcommand\nn         {\nonumber}

\def\mnras{{Mon.~ Not.~ R.~ Astron.~ Soc.~}}

\def\prd{{Phys.~ Rev.~ D.~}}

\def\apj{{Astrophys.~ J.~}}
\def\apjs{{Astrophys.~ J.~ Suppl.~}}
\def\apjl{{Astrophys.~ J.~ Lett.~}}

\def\mnras{{MNRAS}}

\def\prd{{PRD}}

\def\apj{{ApJ}}
\def\apjs{{ApJS}}
\def\apjl{{ApJL}}
\def\aap{{A\&A}}

\newcommand{\be}{\begin{equation}}
\newcommand{\ee}{\end{equation}}
\newcommand{\ba}{\begin{eqnarray}}
\newcommand{\ea}{\end{eqnarray}}

\def\pp1{{\prime}}
\def\pp2{{\prime\prime}}

\def\2D{{\rm 2D}}

\def\Vu{{V_{\mu}}}

\def\bx{{\bf x}}
\def\br{{\bf r}}

\def\bk{{\bf k}}

\def\1Loop{{\rm 1Loop}}

\def\rhob{\bar{\rho}}
\def\Hub{{\rm km}s^{-1}{\rm Mpc}^{-1}}
\def\Msol{h^{-1}M_{\odot}}

\def\Mpc{\, h^{-1}{\rm Mpc}}
\def\Mpccube{\, h^{-3} \, {\rm Mpc}^3}

\def\Gpccube{\, h^{-3} \, {\rm Gpc}^3}
\def\kMpc{\, h \, {\rm Mpc}^{-1}}

\def\dx{{\rm d}^3{\bf x}}
\def\dk{{\rm d}^3{\bf k}}

\def\fun#1#2{\lower3.6pt\vbox{\baselineskip0pt\lineskip.9pt
        \ialign{$\mathsurround=0pt#1\hfill##\hfil$\crcr#2\crcr\sim\crcr}}}

\begin{document}

\title{What do cluster counts really tell us about the Universe?}

\author{Robert~E.~Smith}
\affiliation{Institute for Theoretical Physics, University of Zurich, Zurich CH 8037}
\affiliation{Argelander-Institute for Astronomy, Auf dem H\"ugel 71, D-53121 Bonn, Germany}
\email{res@physik.unizh.ch} 

\author{Laura Marian}
\affiliation{Argelander-Institute for Astronomy, Auf dem H\"ugel 71, D-53121 Bonn, Germany}
\email{lmarian@astro.uni-bonn.de}


\begin{abstract}
  We study the covariance matrix of the cluster mass function in
  cosmology. We adopt a two-line attack: firstly, we employ the
  counts-in-cells framework to derive an analytic expression for the
  covariance of the mass function. Secondly, we use a large ensemble
  of $N$-body simulations in the $\Lambda$CDM framework to test
  this. Our theoretical results show that the covariance can be
  written as the sum of two terms: a Poisson term, which dominates in
  the limit of rare clusters; and a sample variance term, which
  dominates for more abundant clusters. Our expressions are analogous
  to those of \citet{HuKravtsov2003} for multiple cells and a single
  mass tracer. Calculating the covariance depends on: the mass
  function and bias of clusters, and the variance of mass fluctuations
  within the survey volume. The predictions show that there is a
  strong bin-to-bin covariance between measurements. In terms of the
  cross-correlation coefficient, we find $r\gtrsim0.5$ for haloes with
  $M\lesssim 3 \times 10^{14}\Msol$ at $z=0$. Comparison of these
  predictions with estimates from simulations shows excellent
  agreement.  We use the Fisher matrix formalism to explore the
  cosmological information content of the counts. We compare the
  Poisson likelihood model, with the more realistic likelihood model
  of \citet{LimaHu2004}, and all terms entering the Fisher matrices
  are evaluated using the simulations. We find that the Poisson
  approximation should only be used for the rarest objects, $M\gtrsim
  3 \times 10^{14}\Msol$, otherwise the information content of a
  survey of size $V\sim13.5 \Gpccube$ would be overestimated,
  resulting in errors that are $\sim$2 times smaller. As an auxiliary
  result, we show that the bias of clusters, obtained from the
  cluster-mass cross-variance, is linear on scales $>50\Mpc$, whereas
  that obtained from the auto-variance is nonlinear.
\end{abstract}


\maketitle


\section{Introduction}


The last decade of research in cosmology has largely been focused on
devising probes to reveal the physical nature of dark energy and the
origin of the accelerated expansion of the Universe. Among the most
promising probes, as identified for example in \citet{DETF2006}, are
cluster counts.

From a theoretical perspective, the abundance of clusters per unit
solid angle $d\Omega$, is an integral of the mass function over mass
$M$ and volume element $dV$:
\be
\frac{dN}{d\Omega}=\int dz \frac{dV}{d\Omega dz}\int_{M_{\rm th}(z)} dM n(M)\,,
\ee
where $M_{\rm th}(z)$ is a redshift-dependent mass detection threshold
for the clusters and where the mass function is defined as the number
of halos per unit volume and unit mass, i.e. $n(M)=dN/dV/dM$, with $M$
the virial mass. For a wide range of cosmological models, $n(M)$ can
be accurately predicted from the semi-analytical prescriptions based
on the spherical or ellipsoidal collapse model \citep[e.g.][and see
  \S\ref{III4} for more
  details.]{PressSchechter1974,ShethTormen1999}. 

The mass function is primarily sensitive to the statistics of the
initial conditions and to the amplitude $\sigma_8$ and shape of the
matter power spectrum; which in turn depends on the matter density of
the Universe $\Omega_m$, the Hubble parameter $h$, the spectral index
of the primordial power spectrum $n$; and the dark energy equation of
state $w\equiv P_w/\rho_w$, where $P_w$ and $\rho_w$ are the pressure
and energy density of the dark energy.
The volume element integral in the above equation renders the cluster
counts even more sensitive to $\Omega_m$ and $w$.  Measuring the
cluster abundance at different redshifts can constrain a dynamical $w$
and thus enable one to differentiate between a cosmological constant
$\Lambda$ and alternative dark energy scenarios such as quintessence
\citep{WangSteinhardt1998}, or dark energy inhomogeneities coupling to
dark matter \citep{ManeraMota2006}.

For many decades the study of clusters of galaxies has been a
centerpiece for observational cosmology, which has produced many
important results and cosmological inferences. Currently there are
four observational strategies for detecting clusters: $X$-ray
emission \citep[see][and references
  therein]{Borganietal2001,ReiprichBohringer2002,Schueckeretal2003,
  Allenetal2003, Henry2004, Mantzetal2008, Vikhlininetal2009,
  Mantzetal2010a}; optical emission \citep[see][and references
  therein]{Gladdersetal2007, Rozoetal2010}; the Sunyaev--Zel'Dovich
effect \citep[][hereafter SZ effect]{SunyaevZeldovich1972}, i.e. the
up-scattering of CMB photons off hot electrons in the intracluster
medium \citep[see][and references therein]{Vanderlindeetal2010short,
  Planck2011SZ, Muchovejetal2011, Sehgaletal2011short}; weak
gravitational lensing \citep[see][and references
  therein]{Schirmeretal2007, Abateetal2009, Israeletal2010}.

One of the most challenging aspects of deriving cosmological
constraints from cluster counts is the fact that virial masses are not
directly observable: a conversion is needed to translate observables
such as flux, luminosity, temperature and SZ decrement into mass. The
mass-observable relation is degenerate with cosmological parameters,
as shown in \cite{LimaHu2005}, and can severely degrade the inferred
constraints. Substantial progress has been made in calibrating the
mass-observable relation in the recent years, through numerical
simulations, or by comparing different methods against each other
\citep{Zhangetal2007, Zhangetal2008, Okabeetal2010}. \cite{LimaHu2005}
also proposed a self-calibration technique that uses the clustering of
clusters to break the degeneracy between the uncertainties in the
mass-observable relation and cosmological parameters.

Owing to observational challenges, the cluster studies mentioned
earlier employ small numbers of massive clusters (at most a few
hundreds, but in general a few tens) to constrain cosmology. In
obtaining these constraints it is widely assumed that the likelihood
function for the selected clusters follows the Poisson
distribution. Whilst this assumption may be reasonable for the most
massive clusters, $M\sim 10^{15} \Msol$, it will certainly fail at
lower masses.  Future surveys, such as eROSITA
\citep{eROSITA2010short}, \citet{LSST2009}, Euclid \citep{EUCLID2010},
Pan-STARRS\footnote{http://pan-starrs.ifa.hawaii.edu},
\citet{DES2005}, will be able to detect large samples of
intermediate-mass clusters, $M\sim 10^{14} \Msol$. In order to make
accurate inferences from this data, the cluster likelihood function
will require a more complex statistical treatment, and in particular
knowledge about the covariance matrix of the mass function.

This paper is driven by the following two questions: What is the
covariance matrix for measurements of the mass function? How much are
forecasted errors, which rely on the Poisson approximation, affected
by more realistic modelling of the cluster likelihood function? The
main theoretical tools that we shall employ to answer these questions
will be the counts-in-cell formalism introduced by \citet{Peebles1980}
and further developed by \citet[][hereafter HK03]{HuKravtsov2003} and
\citet[][hereafter LH04]{LimaHu2004}.  We shall also compare the
theoretical results obtained via this formalism to measurements
obtained from a large ensemble of $N$-body simulations.

As this paper was nearing submission, a study by
\citet{Valageasetal2011} was reported. This work explores related, but
complimentary, questions to those presented here.

The paper is structured in the following way: in \S\ref{II} we review
the counts-in-cells formalism and also the extension to the cluster
likelihood developed by LH04; in \S\ref{III} we derive the mass
function covariance in a formal way; in \S\ref{IV} we describe the
numerical simulations from which we measure the mass function
covariance; in \S\ref{V} we present a comparison between the measured
and the predicted covariance, and in \S\ref{VI} we use the
Fisher-matrix formalism to estimate the impact that the full
covariance matrix of the mass function has on cosmological
constraints. Finally, in \S\ref{VII} we discuss and summarize our
findings.


\section{Theoretical background}\label{II}


\subsection{The cellular model}\label{II1}


In this section, we give a short description of the counts-in-cell
formalism, used by HK03 to compute the linear-theory sample variance
of cluster counts, and by LH04 to estimate the impact of the latter on
Fisher matrix predictions.

Consider some large cubical patch of the Universe, of volume $\Vu$,
and containing $N$ clusters that possess some distribution of masses.
Let us subdivide this volume into a set of $N_c$ equal cubical cells
and the mass distribution into a set of $N_m$ mass bins. Let the
number of clusters in the $i^{\rm th}$ cell and in the $\alpha^{\rm th}$ mass bin be
denoted $N_{i,\alpha}$. We shall assume that the probability that the
$i^{\rm th}$ cell contains $N_{i, \alpha}$ clusters in the mass bin $\alpha$, is a
Poisson process:
\be P(N_{i, \alpha}|m_{i,\alpha})= \frac{m_{i,\alpha}^{N_{i,
      \alpha}}\exp(-m_{i,\alpha})}{N_{i, \alpha}!} \ .
\label{eq:defPoisson}
\ee
For any quantity $X$, we denote the average over the sampling
distribution--the Poisson process in this case--as $\left<X
\right>_P$, and the ensemble average over many realizations of the
density field as $\overline{X} \equiv \left<X\right>_s$, termed sample
variance in HK03.  The average of $N_{i, \alpha}$ over the sampling
distribution can be written as \citep[see also][]{ColeKaiser1989,MoWhite1996}:
\be
m_{i,\alpha} 
\equiv \overline{m}_{i,\alpha} \left[1+\overline{b}_{\alpha} {\delta}_V(\bx_i) \right], 
\ee
where \mbox{$\overline{m}_{i,\alpha} = \overline{n}_{\alpha} V_i$} is
the ensemble- and Poisson-averaged number of counts in cell $i$ and
mass bin $\alpha$. The volume of the cell and the cell-averaged
overdensity are given by,
\ba 
V_i & = & \int \dx \, W(\bx|\bx_i)\ \ ;\\
{\delta}_V(\bx_i) & = & \frac{1}{V_i}\int \dx \, W(\bx|\bx_i) \delta(\bx) \ .
\ea
where $W(\bx|\bx_i)$ is the window function for the $i$th cell (see
\S\ref{III2} for more details).  The number density and linear bias of
the clusters averaged over the mass bin $\alpha$ are given by:
\ba 
\overline{n}_{\alpha} & = & 
\int_{M_{\alpha}-\Delta M_{\alpha}/2}^{M_{\alpha}+\Delta M_{\alpha}/2} dM n(M) \ ; \\
\overline{b}_{\alpha} & = & 
\frac{1}{\overline{n}_{\alpha}} 
\int_{M_{\alpha}-\Delta M_{\alpha}/2}^{M_{\alpha}+\Delta M_{\alpha}/2} dM b(M) n(M)\ ,
\label{eq:def_bias1}
\ea
where $b(M)$ is the linear bias of haloes of mass $M$.

As was shown in HK03, the correlations in the underlying density field
induce a correlation in the number counts of the cells, defined as:
\ba 
S^{\alpha\beta}_{ij} & \equiv & 
\left<\left(N_{i,\alpha}-\overline{m}_{i,\alpha}\right)
\left(N_{j,\beta}-\overline{m}_{j,\beta}\right)
\right>_{p,s}\ \nn \\
& = & 
\left< 
\left(m_{i,\alpha}-\overline{m}_{i,\alpha}\right)
\left(m_{j,\beta}-\overline{m}_{j,\beta}\right)
\right>_s\nn \\
& = & \overline{m}_{i,\alpha} \overline{m}_{j,\beta} 
\overline{b}_{\alpha}\overline{b}_{\beta} 
\int \frac{\dk}{(2\pi)^3} W^{*}_i(\bk)W_j(\bk) P(k)\ ,
\label{eq:Sdef}\ea
where for independent Poisson processes the probability $P(N_{i,
  \alpha},N_{j, \beta}|m_{i,\alpha},m_{j,\beta})=P(N_{i,
  \alpha}|m_{i,\alpha})P(N_{j, \beta}|m_{j,\beta})$. In the last line
we introduced the power spectrum $P(k)$ as the Fourier transform of
the correlation function $\xi$,
\be 
\xi(\br) \equiv \left<\delta(\bx_i)\delta(\bx_j)\right>_{s} 
= \int \frac{\dk}{(2\pi)^3} P(k) \exp\left(-i\bk\cdot\br\right). 
\ee
$W_i(\bk)$ is the Fourier transform of the cell window function and
$\br=\bx_i-\bx_j$ (see \S\ref{III2}).


\subsection{The Gauss-Poisson likelihood function for counts in cells}
\label{II2}


The likelihood of drawing a particular set of cluster counts
\mbox{${\bf
    N}\in\{N_{1,1},\dots,N_{N_c,1},N_{1,2},\dots,N_{N_c,N_m}\}$} in
the cells, given a model for the counts in the cells
\mbox{$\overline{\bf
    m}\in\{\overline{m}_{1,1},\dots,\overline{m}_{N_c,1},
  \overline{m}_{N_c,2}\dots,\overline{m}_{N_c,N_m}\}$} was written by
LH04:
\be 
{\mathcal L}({\bf N}|\overline{\bf m},{\bf S}) = \int d^{{\mathcal N}}\!m
\left[\prod_{\alpha=1}^{N_m}\prod_{i=1}^{N_c} P(N_{i,\alpha}|m_{i,\alpha}) \right] 
G({\bf m}|\overline{\bf m},{\bf S}) 
\label{eq:likelihood1}
\ee
with ${\mathcal N}=N_c\times N_m$, and where it was assumed that the
statistics of the cell-averaged density field are described by a
multivariate Gaussian:
\be G({\bf m}|\overline{\bf
  m},{\bf S}) \equiv \frac{(2\pi)^{-N/2}}{|S|^{1/2}} \exp\left[-\frac{1}{2}
({\bf m}-\overline{\bf m})^{T}{\bf S}^{-1}({\bf m}-\overline{\bf m})\right] 
\ , 
\label{eq:likelihoodG}
\ee
%
with ${\bf S}$ defined in \Eqn{eq:Sdef}. Measurements of the
bispectrum of the CMB have shown that the statistics of the initial
fluctuations are very nearly Gaussian \citep{Komatsuetal2010short}.
Whilst we know that nonlinear growth of structure in the present epoch
drives the statistics of the density field to become non-Gaussian, in
the limit that the cells are large compared to the coherence length of
the field, we expect that the Gaussian approximation will be very
good.

At this point we may also be more precise about what we mean by
ensemble and Poisson averages:
\be \left<X({\bf N})\right>_{P,s} \equiv \sum_{N_{1,1}=0}^{\infty}
\dots \sum_{N_{N_c,N_m}=0}^{\infty} {\mathcal L}({\bf N}|\overline{\bf
  m},{\bf S}) X({\bf N}) \ .
\label{eq:formal_average}
\ee
Equation~(\ref{eq:likelihood1}) can be simplified in two limits:
\begin{itemize}
\item{\bf Case I}: In the limit that the ensemble average
variance is much smaller than the Poisson variance: i.e. $S_{ii}\ll
\overline{m}_{i}$. In this case, the Gaussian effectively becomes a
delta function centred on $\overline{{\bf m}}$ and the likelihood
simply becomes a product of Poisson probabilities:
\be 
{\mathcal L}({\bf N}|\overline{\bf m})\approx 
\prod_{\alpha=1}^{N_m}\prod_{i=1}^{N_c} P(N_{i,\alpha}|\overline{m}_{i,\alpha})
\label{eq:likelihoodP}\ .
\ee
\item {\bf Case II}: In the limit that the number of counts in
each cell and mass bin is large, then the Poisson process becomes a
Gaussian:
\be \prod_{\alpha=1}^{N_m}\prod_{i=1}^{N_c} P(N_{i,\alpha}|m_{i,\alpha})
\approx G({\bf N}|{\bf m},{\bf M})\ , \label{eq:likelihoodG2}
\ee
where ${\bf M}\rightarrow
M^{ij}_{\alpha\beta}=\delta^{K}_{i,j}\delta^{K}_{\alpha,\beta}
m_{i,\alpha}$. Hence, as shown in LH04, the likelihood function
becomes,
\be {\mathcal L}({\bf N}|\overline{m},{\bf S}) \approx \int d^{{\mathcal N}}\!m
G({\bf N}|{\bf m},{\bf M}) G({\bf m}|\overline{\bf m},{\bf S}) \ee
and via the convolution theorem this can be approximated as a Gaussian
with shifted mean and augmented covariance matrix:
\be {\mathcal L}({\bf N}|\overline{\bf m},{\bf S}) \approx 
G({\bf N}|\overline{\bf m},{\bf C}) \ \ ;\ \ {\bf C}=\overline{{\bf M}}+{\bf S}
\label{eq:likelihoodApprox}\ ,\ee
where $\overline{\bf
  M}\rightarrow\overline{M}^{ij}_{\alpha\beta}=\delta^{K}_{i,j}\delta^{K}_{\alpha,\beta}
\overline{m}_{i,\alpha}$. Note that in the above equation, the
approximate sign is used since negative number counts are formally
forbidden \citep[for a more detailed discussion of this
  see][]{HuCohn2006}.
\end{itemize}


\section{Covariance of the mass function}\label{III}


The final result of \S\ref{II} is that in the limit of a large number
of counts per cell, the joint likelihood for all the cells is a
Gaussian with model mean $\overline{\bf m}$ and with a covariance
matrix, ${\bf C}=\overline{{\bf M}}+{\bf S}$. In the following
section, we shall use these results to answer the question: What is
the covariance matrix for measurements of the mass function?


\subsection{A formal approach}\label{III1}

The mass function $n(M)$ is the number density of clusters in a volume
$V$, per unit mass. Using our counts in cells distribution, an
estimator for the mass function in the $i^{\rm th}$ cell is,
\be \hat{n}_i(M_{\alpha}) = \frac{N_{i,\alpha}}{V_i \Delta M_{\alpha}} \ ,\ee
which, if we average over all cells and all cells have equal volume,
becomes
\be \hat{n}(M_{\alpha}) = \frac{1}{\Vu \Delta M_{\alpha}}\sum_i N_{i,\alpha}.
\ee
The above estimate is unbiased, and its expectation value
$\overline{n}(M_\alpha) \equiv \left<\hat{n}(M_{\alpha})\right>_{P,s}
$ can be formally calculated using Eq.(\ref{eq:formal_average}): 
\ba 
\overline{n}(M_\alpha) & = & \sum_{N_{1,1}=0}^{\infty}\dots \sum_{N_{N_c,N_m}=0}^{\infty}
{\mathcal L}({\bf N}|\overline{\bf m},{\bf S}) 
\sum_i \frac{N_{i,\alpha}}{\Vu \Delta M_{\alpha}}\nn \\
& = &
\int d^{{\mathcal N}}\!m G({\bf m}|\overline{\bf m},{\bf S})
\sum_{N_{1,1}=0}^{\infty}P(N_{1,1}|m_{1,1}) \dots  \nn \\
& \times & 
\sum_{N_{N_c,N_m}=0}^{\infty}P(N_{N_c,N_m}|m_{N_c,N_m}) \sum_{i=1}^{N_c} 
\frac{N_{i,\alpha}}{\Vu \Delta M_{\alpha}} \nn \\
& = & \frac{1}{\Vu \Delta M_{\alpha}}\int d^{{\mathcal N}}\!m G({\bf m}|\overline{\bf m},{\bf S})
\sum_{i} m_{i,\alpha} \nn \\
& = & 
\sum_{i=1}^{N_c}\frac{\overline{m}_{i,\alpha}}{\Vu \Delta M_{\alpha}}
\ea
In a similar fashion, the covariance matrix of the cluster mass
function can also be calculated:
\ba 
{\mathcal M}_{\alpha\beta} & \equiv & \left<
\left[n(M_{\alpha})-\overline{n}(M_{\alpha})\right]
\left[n(M_{\beta})-\overline{n}(M_{\beta})\right]
\right>_{s,P} \nn \\
& = & 
\sum_{i,j}\frac{\left<N_{i,\alpha}N_{j,\beta}\right>_{s,P}}{\Vu^2 \Delta M_{\alpha}\Delta M_{\beta}}
-\overline{n}(M_{\alpha})\overline{n}(M_{\beta})\label{eq:massfuncov1},
\ea
where the expectation of the product of the counts can be written
\ba 
\sum_{i,j}\left<N_{i,\alpha}N_{j,\beta}\right>_{s,P} \!\!\! & = \!\!\! & \!\!\!
\sum_{N_{1,1}=0}^{\infty}\dots \!\!\!\sum_{N_{N_c,N_m}=0}^{\infty}
\!\!\!{\mathcal L}({\bf N}|\overline{\bf m},{\bf S}) 
\sum_{i,j} N_{i,\alpha}N_{j,\beta} \nn\\
& & \hspace{-2.5cm} =
\int d^{{\mathcal N}}\!m G({\bf m}|\overline{\bf m},{\bf S}) \left[\sum_{i,j,i\ne j\cup \alpha\ne\beta} m_{i,\alpha}m_{j,\beta}+
\sum_{i} \left<N^2_{i,\alpha}\right>\right] \ .
\nn \\
\ea
Recall that \mbox{$\overline{m}_{i,\alpha} =
  \overline{n}(M_{\alpha})\Delta M_{\alpha} V_i$} and that for the
Poisson distribution we have:
\mbox{$\left<X^2\right>=\left<X\right>[1+\left<X\right>]$}. On
inserting these relations into the above equation, and on completing
the sums, we find:
\ba 
\sum_{i,j}\left<N_{i,\alpha}N_{j,\beta}\right>_{s,P}\!\!\! & = &
\!\!\!\int d^{{\mathcal N}}\!m G({\bf m}|\overline{\bf m},{\bf S}) \nn \\
&   & \hspace{-1cm} \times \sum_{i,j} \left[m_{i,\alpha}m_{j,\beta} +
m_{i,\alpha}\delta^{K}_{i,j} \delta^{K}_{\alpha,\beta}\right]  \nn \\
& & \hspace{-1cm} = 
\sum_{ij}\left[S_{ij}^{\alpha\beta}+\overline{m}_{i,\alpha}\overline{m}_{j,\beta}
+\overline{m}_{i,\alpha} \delta^{K}_{i,j} \delta^{K}_{\alpha,\beta}\right],
\ea
where in the last line we used \Eqn{eq:Sdef}. On inserting this result
back into \Eqn{eq:massfuncov1}, we obtain
\ba 
{\mathcal M}_{\alpha\beta} & = &
\sum_{ij}\frac{\left[\overline{m}_{i,\alpha}
\delta^{K}_{i,j} \delta^{K}_{\alpha,\beta}+S_{ij}^{\alpha\beta}\right]}{\Vu^2 \Delta M_{\alpha}\Delta M_{\beta}} \nn
\\
& = & 
\frac{\delta^{K}_{\alpha,\beta} \overline{n}(M_{\alpha})}{\Vu \Delta M_{\alpha}} + \frac{\overline{n}(M_{\alpha}) \overline{n}(M_{\beta})\overline{b}_{\alpha}\overline{b}_{\beta}}{\Vu^2} \nn \\
& & \times \sum_{ij}V_i V_j
\int \frac{\dk}{(2\pi)^3} W^{*}_i(\bk)W_j(\bk) P(k) \ .
\label{eq:massfuncov2}
\ea
Considering the first term in the above, we may simplify this
expression by performing the sums over $i$ and $j$, and the window
functions i.e.
\ba 
\sum_{i}V_i W_i(\bk) & = &\sum_{i}V_i \int \dx 
\exp\left[i\bk\cdot\bx\right] W(\bx|\bx_i) \nn \\
 & = & \int \dx \exp\left[i\bk\cdot\bx\right]\sum_{i}V_i W(\bx|\bx_i) 
= \Vu \widetilde{W}(\bk).\nn\\
\ea
Hence, we have that the covariance matrix can be written:
\be
{\mathcal M}_{\alpha\beta} = \overline{n}(M_{\alpha})\overline{n}(M_{\beta})
\overline{b}_{\alpha}\overline{b}_{\beta}
\sigma^2(\Vu)+\frac{\delta^{K}_{\alpha,\beta}\overline{n}(M_{\alpha})}{\Vu \Delta M_{\alpha}}
\label{eq:massfuncov3},
\ee
where $\sigma^2(\Vu)$ is the mass density variance in the entire volume
\be \sigma^2(\Vu) \equiv \int \frac{\dk}{(2\pi)^3}
\left|\widetilde{W}(\bk)\right|^2 P(k) \ .\ee
From \Eqn{eq:massfuncov3} it can be seen that the crucial quantity
which controls the covariance between estimates of the mass function
in different mass bins is $ \sigma(\Vu)$. The strength of the
covariance is also modulated by the linear bias and the mass function
in each of the bins considered.


\subsection{A short-cut to the covariance}\label{III2}


Whilst in the above we have presented a formal derivation of the mass
function covariance from the HK03 and LH04 formalism, there is a more
intuitive approach to arriving at the same result as given by
\Eqn{eq:massfuncov3}, which we now mention.

Let us consider the limiting case where we have a single cell that
fills the whole of our sample space $V_i\rightarrow \Vu$; also
$m_{i,\alpha}\rightarrow m_{\alpha}$ and similar for all the other
quantities defined in the cells. The above formalism still applies,
and we have that the covariance matrix of mass function can be
written:
\ba
{\mathcal M}_{\alpha\beta} & = &
\frac{S_{ii}^{\alpha,\beta} }{V_{\mu}^2\Delta M_{\alpha}\Delta M_{\beta}} 
+ \delta^{K}_{\alpha,\beta} \frac{m_{i,\alpha}}{V_{\mu}^2\Delta M_{\alpha}\Delta M_{\beta}} \nn \\
& = & \overline{n}(M_{\alpha}) \overline{n}(M_{\beta})\overline{b}_{\alpha}\overline{b}_{\beta}
\sigma^2(\Vu)  + \delta^{K}_{\alpha,\beta} \frac{n(M_{\alpha})}{V_{\mu}\Delta M_{\alpha}} \ ,
\ea
where $\sigma(\Vu)$ is the variance in the total volume.


\subsection{The cross-correlation coefficient}\label{III3}


As a direct corollary to the previous results, we may write an
expression for the correlation matrix, which is defined
\be 
r_{\alpha\beta} \equiv \frac{{\mathcal M}^{\alpha\beta}}
{\sqrt{{\mathcal M}^{\alpha\alpha} {\mathcal M}^{\beta\beta}}} \ .
\ee
On factoring out $[\overline{n}(M_{\alpha})/\Vu\Delta
  M_{\alpha}]^{1/2}$ from $\sqrt{{\mathcal M}^{\alpha\alpha}}$ in the
denominator, and a similar term from $\sqrt{{\mathcal
    M}^{\beta\beta}}$, and on using the fact that
$\overline{m}_{\alpha}= \overline{n}(M_{\alpha})\Delta M_{\alpha}
\Vu$, we find
\be 
r_{\alpha\beta} =
\frac{\sqrt{\overline{m}_{\alpha}\overline{m}_{\beta}} 
\,\overline{b}_{\alpha}\overline{b}_{\beta}\sigma^2(\Vu)
+\delta^{K}_{\alpha\beta}}
{
\left[1+\overline{m}_{\alpha}\overline{b}^2_{\alpha}\sigma^2(\Vu)\right]^{1/2}
\left[1+\overline{m}_{\beta}\overline{b}^2_{\beta}\sigma^2(\Vu)\right]^{1/2}}\ .
\label{eq:corrcoef}
\ee
Two limits are apparent: when
$\sqrt{\overline{m}_{\alpha}\overline{m}_{\beta}}\,
\overline{b}_{\alpha}\overline{b}_{\beta}\sigma^2(\Vu)\ll1$, then
$r_{\alpha\beta}\rightarrow\delta^{K}_{\alpha,\beta}$ and the mass
function covariance matrix is decorrelated; this would happen for the
case of rare halos, for which the mass function is very small. On the
other hand, when $\sqrt{\overline{m}_{\alpha}\overline{m}_{\beta}}\,
\overline{b}_{\alpha}\overline{b}_{\beta}\sigma^2(\Vu)\gg1$, then
$r_{\alpha\beta}\rightarrow 1$ and the covariance matrix is fully
correlated. This would be the case for smaller halos, for which the
mass function is quite large. 

Finally, we note that taking $\Vu\rightarrow\infty$ and hence
$\sigma(\Vu)\rightarrow 0$, \emph{does not} guarantee that the
correlation between different mass bins is negligible. As the above
clearly shows, it is the quantity $\Vu\sigma^2(\Vu)$ that is required
to vanish for negligible correlation to occur. For a power-law power
spectrum, we would have that $\Vu\sigma^2(\Vu)\propto R^3
R^{-(3+n)}\propto R^{-n}$, which only vanishes for $n>0$. For CDM we
have a rolling spectral index, and $n>0$ for $k\lesssim0.01\kMpc$,
which implies that $L_{\rm box}\gtrsim 500\Mpc$ for the covariance to
diminish.

\vspace{0.2cm}

At this juncture, we point out that \Eqns{eq:massfuncov3}{eq:corrcoef}
constitute the main analytic results of this work, and all which
follows will be concerned with their validation and implications.

%

\begin{table*}
\caption{{\tt zHORIZON} cosmological parameters. Columns
are: density parameters for matter, dark energy and baryons; the
equation of state parameter for the dark energy $w$;
normalization and primordial spectral index of the power spectrum;
dimensionless Hubble parameter.
\label{tab:zHORIZONcospar}}
\vspace{0.2cm}
\centering{
\begin{tabular}{c|ccccccc}
\hline 
Cosmological parameters & $\Omega_m$ & $\Omega_{DE}$ & $\Omega_b$ & $w$  &  
$\sigma_8$  & $n$ &  $H_{0} [\Hub]$ \\
\hline
{\tt zHORIZON-I      }  & 0.25\ &  0.75 & 0.04 &  -1  &  0.8  & 1.0 & 70.0\\
{\tt zHORIZON-V1a/V1b}  & 0.25\ &  0.75 & 0.04 &  -1  &  0.8  & 0.95/1.05 & 70.0\\
{\tt zHORIZON-V2a/V2b}  & 0.25\ &  0.75 & 0.04 &  -1  &  0.7/0.9  & 1.0 & 70.0\\
{\tt zHORIZON-V3a/V3b}  & 0.2/0.3\ &  0.7 & 0.04 &  -1  &  0.8  & 1.0 & 70.0\\
{\tt zHORIZON-V4a/V4b}  & 0.25\ &  0.8 & 0.04 &  -1.2/-0.8  &  0.8  & 1.0 & 70.0\\
\end{tabular}}
\end{table*}


\begin{table*}
\caption{\small {\tt zHORIZON} numerical parameters. Columns are: number of
particles, box size, particle mass, force softening, number of realizations, 
and total simulated  volume. 
\label{tab:zHORIZONsimpar}}
\vspace{0.2cm}
\centering{
\begin{tabular}{c|cccccc}
\hline
Simulation Parameters & 
$N_{\rm part}$ & $L_{\rm sim}\,[{\rm Mpc}\,h^{-1}]$ & $m_p [\Msol]$ &
 {$l_{\rm soft}\,[{\rm kpc}\,h^{-1}]$} & $N_{\rm ensemb}$ & $V_{\rm tot}[\Gpccube]$\\
\hline
{\tt zHORIZON-I}  & $750^3$ & 1500 & $5.55\times 10^{11}$ & 60 & 40 & 135 \\
{\tt zHORIZON-V1, -V2, -V4} & $750^3$ & 1500 & $5.55\times 10^{11}$ & 60 & 4  &  13.5 \\
{\tt zHORIZON-V3a} & $750^3$ & 1500 & $4.44\times 10^{11}$ & 60 & 4  &  13.5 \\
{\tt zHORIZON-V3b} & $750^3$ & 1500 & $6.66\times 10^{11}$ & 60 & 4  &  13.5\\
%
\end{tabular}}
\end{table*}


\subsection{Ingredients for evaluating the covariance}
\label{III4}

To evaluate the covariance matrix we need to provide models for
$n(M)$, $b(M)$ and the Fourier transform of the survey window
function.

To compute $n(M)$ and $b(M)$ we employ the mass function and bias
models presented in \citet{ShethTormen1999}:
\be \frac{dn}{d\log M}=\frac{\rhob}{M} f_{\rm ST}(\nu) \frac{d\log\nu}{d \log M}\ ;\ee
%
%
\be 
f_{\rm ST}(\nu)=A\sqrt{\frac{2q}{\pi}} \nu \left[
  1+(q \nu^2)^{-p}\right]\exp\left[-\frac{q\nu^2}{2}\right] \ ;
\label{eq:ST_mf}
\ee
\be
b_{\rm ST}(\nu)=1+\frac{q\nu^2-1}{\delta_{\rm sc}} 
+ \frac{2p/\delta_{\rm sc}}{1+(q \nu^2)^p}\ ,
\label{eq:ST_bias}
\ee
where $A=0.3222,\, q=0.707,\,p=0.3$. In the above we have introduced
the peak-height \mbox{$\nu(M)\equiv \delta_{\rm sc}/\sigma(M)$}, where
$\delta_{\rm sc}=1.686/D(z)$ is the spherical overdensity for
collapse, and where $\sigma^2(M)$ is the variance of the linear
density field extrapolated to $z=0$, smoothed with a spherical top-hat
filter of radius $R$ (see below for more details). This radius is
defined so as to enclose a mass $M=4\pi\overline{\rho}R^3/3$, with
$\overline{\rho}$ the mean matter density of the Universe at the present epoch.

For the survey window function we shall consider two simple
examples. The first is a cubical top-hat, defined by:
\be
W(\bx|\bx_j) = 
\left\{ 
\begin{array}{lc}
1/V_j\,, &  x^l_j-L_{\rm box}/2\leq x^l < x_j^l+L_{\rm box}/2 \nn \\
0, & \mbox{otherwise} 
\end{array}
\right. \ , \label{eq:defTHC}
\ee
where $l\in\{1,2,3\}$ denotes the Cartesian components of the vectors,
$j$ is the cell index, and $L_{\rm box}$ is the size of the cell of volume
$V_j=L_{\rm box}^3$. The Fourier transform of this top-hat window function is:
\be
W_j(\bk) = \exp(i \bk \cdot \bx_{j})  \prod_{l=1}^{3} 
j_0\left(\frac{k_l L_{\rm box}}{2}\right),
\label{eq:windowFTC}
\ee
where $j_0(y)\equiv \sin y/y$ is the zeroth order spherical Bessel
function. The volume variance for this window function is:
\ba 
\sigma^2(\Vu) & = & \prod_{l=1}^{3} \left\{ \int_{-\infty}^{\infty}  
\frac{dk_{l}}{2\pi} \right\}
 P(k_1,k_2,k_3)\,|\widetilde{W}(\bk)|^2,\nn \\
              & = & 8 \prod_{l=1}^{3} \left\{ \int_{k_{\rm min}}^{k_{\rm max}}
  \frac{dk_{l}}{2\pi} \right\}
 P(k_1,k_2,k_3)\,|\widetilde{W}(\bk)|^2,
\label{eq:sigmacube}
\ea
where in the second equality we have used the isotropy of the power
spectrum, e.g. $P(k_1,k_2,k_3)=P(-k_1,k_2,k_3)$. In \Eqn{eq:sigmacube}
we use the following relation:
\be
|\widetilde{W}(\bk)|^2 = \prod_{l=1}^{3} j_0^2\left(\frac{k_l
  L_{\rm box}}{2}\right). 
\ee

The second window function is a spherical top-hat:
\ba
W_j(r) = \left\{ \begin{array}{lc}
3/(4\pi R^3),& |\bx_j| < r < |\bx_j| + R \nn \\
  0, & \mbox{otherwise}\,,
\end{array} \right.
\label{eq:defTHS}
\ea
where $R$ is the radius of the spherical top-hat. The variance of the
density field in this case has the familiar form:
\be \sigma^2(\Vu) = \frac{1}{2\pi^2}\int_{k_{\rm min}}^{k_{\rm max}} dk\,
k^2 P(k)\widetilde{W}^2(kR).
\label{eq:SigmaV0STH}
\ee for which the Fourier transform is given by:
\be
\widetilde{W}(x)=\frac{3}{x^3}[\sin x -x\cos x ]\ ; \ \ x\equiv kR\ .
\label{eq:windowFTS}
\ee

On a technical note, we point out that for the $k$-space integrals
given by \Eqns{eq:sigmacube}{eq:SigmaV0STH}, we have introduced lower
and upper limits $k_{\rm min}>0$ and $k_{\rm max}$, respectively. For
a real survey, the upper limit is decided by the resolution of the
instrument used. If the measurements are made from numerical
simulations, which is the case with this work, the softening length of
the simulations will dictate the largest frequency Fourier mode
available: $k_{\rm max}=2\pi/l_{\rm soft}$ and for our simulations
$k_{\rm max}\sim100\kMpc$. However, in practice the largest useful
Fourier mode is much smaller, and occurs where the shot-noise
correction to the power spectrum becomes comparable with the signal
\citep{Smithetal2003}.

The lower limit $k_{\rm min}$ is a more complex issue. In the case of
simulations, no modes with wavelength larger than the simulation box
$L_{\rm sim}$ can contribute to the variance, which suggests the
straightforward solution of adopting $k_{\rm min}=2\pi/L_{\rm sim}$.
Since we are attempting to confront the theory with the {\em reality}
defined by simulations, we shall always assume this cut-off
scale. However, for real surveys, the variance on a given scale will
be affected by the existence of modes on scales larger than the size
of the survey.  We therefore recommend in this case $k_{\rm
  min}\rightarrow 0$, or at least the inverse horizon size at the
redshift of the survey. For more discussion of the importance of
$k_{\rm min}$ for the predictions of the variance, see discussion in
Appendix~\ref{AI}.

Note that in the above we shall relate the radius $R$ of the spherical
top-hat to that of the cubical top-hat function, through the relation
$R=\left(3/4\pi\right)^{1/3} L_{\rm box}$. In other words the volumes
of the spherical and cubical sample volumes are taken to be identical.


\section{$N$-body simulations}\label{IV}


We study the covariance matrix with a suite of 40 large numerical
simulations, executed on the zBOX-2 and \mbox{zBOX-3} supercomputers
at the Institute for Theoretical Physics, University of
Z\"{u}rich. For all realizations snapshots were output at: $z=\{5, 4,
3, 2, 1, 0.5, 0\}$. We shall refer to these simulations as the {\tt
  zHORIZON} Simulations. 

Each of the {\tt zHORIZON} simulations was performed using the
publicly available {\tt Gadget-2} code \citep{Springel2005}, and
followed the nonlinear evolution under gravity of $N=750^3$ equal-mass
particles in a comoving cube of length $L_{\rm sim}=1500\Mpc$. The
cosmological model is similar to that determined by the WMAP
experiment \citep{Komatsuetal2009}. We refer to this cosmology as the
fiducial model. The transfer function for the simulations was
generated using the publicly available {\tt cmbfast} code
\citep{SeljakZaldarriaga1996,Seljaketal2003b}, with high sampling of
the spatial frequencies on large scales. Initial conditions were set
at redshift $z=50$ using the serial version of the publicly available
{\tt 2LPT} code
\citep{Scoccimarro1998,Crocceetal2006}. Table~\ref{tab:zHORIZONcospar}
summarizes the cosmological parameters that we simulate and
Table~\ref{tab:zHORIZONsimpar} summarizes the numerical parameters
used.

In this paper we also study the Fisher matrix of cluster counts for
which we use another series of simulations. Each of the new set is
identical in every way to the fiducial model, except that we have
varied one of the cosmological parameters by a small amount. For each
new set we have generated 4 simulations, matching the random
realization of the initial Gaussian field with the corresponding one
from the fiducial model. The four parameter variations that we
consider are $\{n\rightarrow\{0.95, 1.05\},\,\sigma_8\rightarrow\{0.7,
0.9\},\,\Omega_m\rightarrow \{0.2, 0.3\},\, w\rightarrow\{-1.2,
-0.8\}\}$, and we refer to each of the sets as {\tt
  zHORIZON-V1a,b},\dots,{\tt zHORIZON-V4a,b}, respectively. Again, the  
full details are summarized in Tables \ref{tab:zHORIZONcospar} \&
\ref{tab:zHORIZONsimpar}.

Lastly, dark matter halo catalogues were generated for all snapshots
of each simulation using the Friends-of-Friends (FoF) algorithm
\citep{Davisetal1985}, with the standard linking-length parameter
$b=0.2$, where $b$ is the fraction of the inter-particle spacing. For this
we employed the fast parallel {\tt B-FoF} code, kindly provided to us
by V.~Springel. The minimum number of particles for which an object is
considered to be a bound halo was set at 20 particles. This gave a
minimum host halo mass of $M\sim10^{13} M_{\odot}/h$.


\begin{figure}
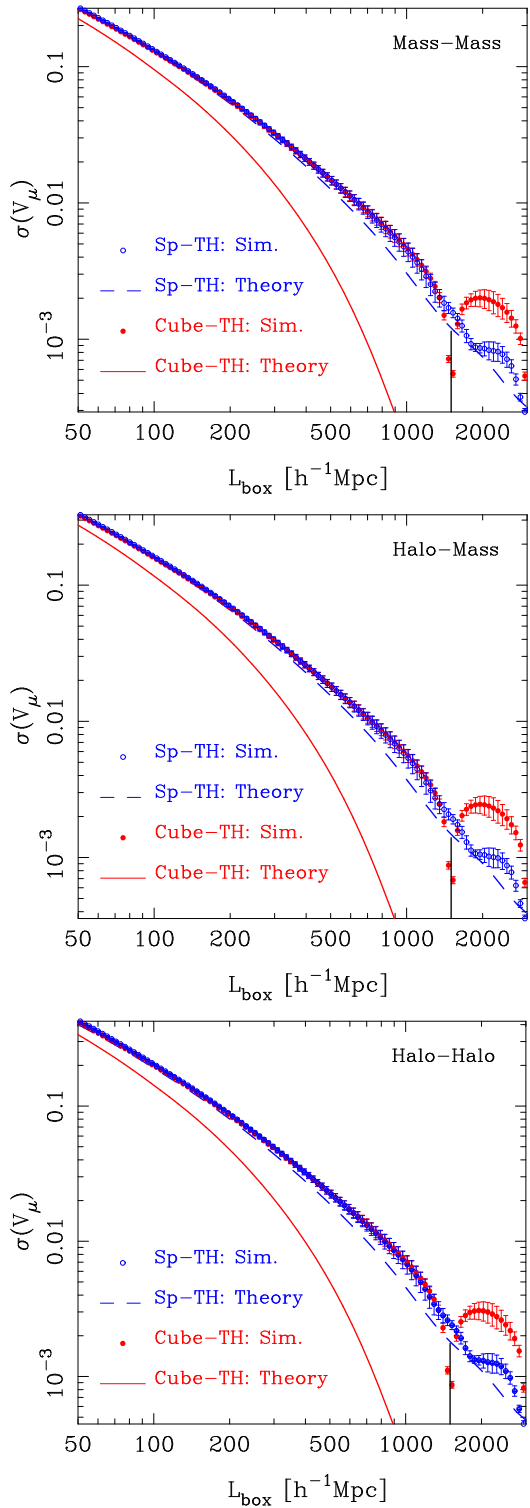

\centering{
  \includegraphics[width=6.9cm]{FIGS/SigVvsV.SigmaSimPlot.MM.ps}}\vspace{0.2cm}
\centering{
  \includegraphics[width=6.9cm]{FIGS/SigVvsV.SigmaSimPlot.HM.ps}}\vspace{0.2cm}
\centering{
  \includegraphics[width=6.9cm]{FIGS/SigVvsV.SigmaSimPlot.HH.ps}}\vspace{0.2cm}
\caption{\small{The r.m.s. density variance as a function of the
    sample volume size $L_{\rm box}$.  From top to bottom, we show
    results for $\sigma_{\rm mm}(\Vu)$, $\sigma_{\rm hm}(\Vu)$,
    $\sigma_{\rm hh}(\Vu)$, respectively. In each panel, blue empty
    and solid red circles denote measurements from the simulations,
    made using the spherical and cubical top-hat filter functions.
    The corresponding analytical predictions for the variance are
    denoted by the dashed blue and solid red lines, respectively. The
    size of the simulation box $1500 \Mpc$ is indicated by a black
    vertical lines, and the measurements are an average of 40
    simulations. }
\label{fig:sigma_multiple}}
\end{figure}


\section{Results}\label{V}


In this section we confront the counts-in-cells theory with the
results from $N$-body simulations.


\subsection{Cell variance in simulations and theory}\label{V1}


Since $\sigma^2(\Vu)$ plays a vital role in determining the strength
of any covariance in the mass function measurements, we shall make a
detailed study of it, for both window functions discussed in
\S\ref{III} and considering volumes of varying size. We evaluate
$\sigma^2(\Vu)$ in two different ways, analytically and from $N$-body
simulations. Furthermore, owing to concerns regarding the impact of
nonlinear bias and mass evolution, we also compute the matter-matter,
halo-matter, and halo-halo variance, which we denote as $\sigma^2_{\rm
  mm}(\Vu)$, $\sigma^2_{\rm hm}(\Vu)$ and $\sigma^2_{\rm hh}(\Vu)$,
respectively. Comparing these quantities will then make clear any
departures from linearity.

Our analytical approach to determining the variances is based on
standard quadrature routines to evaluate the theoretical expressions:
for \Eqn{eq:sigmacube}, we use the multi-dimensional Monte-Carlo
integration routine {\tt VEGAS}; and for \Eqn{eq:SigmaV0STH}, we use
the {\tt QROMB} routine \citep[for more details
  see][]{Pressetal1992}. In evaluating these integrals we take the
linear theory power spectrum matching our simulations, fully described
in \S\ref{IV}. Also, we take the largest mode in the simulation box to
determine the lower limit of the $k$-integrals $k_{\rm min}$.

The second method is one of brute force: we measure $\sigma^2_{\rm
  mm}(\Vu)$, $\sigma^2_{\rm hm}(\Vu)$ and $\sigma^2_{\rm hh}(\Vu)$
directly from the ensemble of simulations. Our estimator for the
variances can be expressed as:
\ba \hat{\sigma}^2_{\rm ab} & \equiv & \int
\frac{\dk}{(2\pi)^3}P_{ab}(k)W^2(kL_{\rm box}) \nn\\ 
& \approx & \frac{1}{\Vu}
\sum_{i,j,k=-N_{\rm g}/2+1}^{N_{\rm g}/2} \hat{P}_{\rm ab}(\bk_{ijk})
|W(k_{ijk},L_{\rm box})|^2, \label{eq:sigmalattice}
\ea
where the indices $(i,j,k)$ label the Fourier mesh cell and $k_{ijk}$
the magnitude of the wavenumber corresponding to that cell. The total
number of grid cells considered is $N_{\rm g}^3$; also, $a$ and $b$
are $\in\{\rm m, h\}$, and $\hat{P}_{ab}(\bk_{ijk})\equiv \Vu
\delta_{a}(\bk_{ijk})^{*}\delta_{b}(\bk_{ijk})$ are estimates of the
various auto- and cross-power spectra. The window functions are as
given in \S\ref{III}. The estimates of the variance also require a
correction for shot-noise, which for the halo-halo variance we
implement in the following way:
\be 
\hat{\sigma}^2_{\rm hh,c} =  \hat{\sigma}^2_{\rm hh,d}-\frac{1}{N_h}
\sum_{i,j,k}^{} |W(k_{ijk},L_{\rm box})|^2,
\ee
where $N_h$ is the number of halos in the considered mass bin, and
$\hat{\sigma}^2_{\rm hh,c}$ and $\hat{\sigma}^2_{\rm hh,d}$ are the
variance of the continuous and discrete halo density fields,
respectively. There is a similar shot-noise correction for the
matter-matter variance; we assume that the halo-mass cross-variance
requires no such correction. Note that the above method for estimating
$\sigma(\Vu)$ is not the conventional one, where one partitions the
real space counts into cells and then computes the variance of that
distribution. However, it should be entirely equivalent, but with the
added advantages of being fast, since we are using an FFT, and
allowing for the computation of the variance in arbitrary cell
structures.

Rather than testing all of the halo mass bins that we will employ
later for the mass function covariance, we have chosen to show results
for all the haloes in the simulation with $M>10^{13}\Msol$.
Figure~\ref{fig:sigma_multiple} presents our results for $\sigma_{\rm
  mm}(\Vu)$, $\sigma_{\rm hm}(\Vu)$ and $\sigma_{\rm hh}(\Vu)$ as a
function of the cubical window function size, $L_{\rm box}$; recall
that for the spherical window we take the radius to be
$R=\left(3/4\pi\right)^{1/3} L_{\rm box}$. In all three panels, the
points represent results from the $N$-body simulations, whereas the
lines denote the analytical integrals. The red full circles and solid
lines are obtained by smoothing the density field with the cubical
top-hat, while the blue empty circles and dashed lines denote
smoothing with the spherical top-hat function. The simulation results
represent the mean of the 40 realizations, with errors appropriate for
a single run. The size of the simulation box ($L_{\rm sim}=1500 \Mpc$)
is marked through a vertical black line on the horizontal axis. The
effects of the shot-noise corrections on the estimates of
$\hat{\sigma}^2_{\rm mm,c}$ and $\hat{\sigma}^2_{\rm hh,c}$ are too
small to be noticed on this log--log plot.

As expected for a hierarchical mass distribution, in all cases the
variance decreases steeply with the increasing box size. On comparing
the results obtained from the simulations for the two window
functions, we find very good agreement up until the size of
the cubical region becomes similar to the size of the simulation cube.
At this scale, the variance from the cubical window function displays
a significant loss in signal. For scales larger than the simulation
box, the smoothing result become somewhat meaningless and unstable due
to the oscillatory nature of both window functions, which can be seen
from the measurements. 

Turning to the evaluation of the theoretical expressions for the
variance, we see that, in the case of the spherical top-hat there is
excellent agreement between the simulations and the theory on small
scales, $L_{\rm box}<200\Mpc$. For $L_{\rm box} \geq 200 \Mpc$, the
linear expressions underestimate the measurements by $\approx 20 \%$
or even more. However, on comparing the theoretical predictions for
the cubical filter function with the measurements, we find a large
discrepancy. We tested whether this was due to an error in the {\tt
  VEGAS} evaluation of the integrals. An independent check with {\tt
  mathematica} produced the same results. 

After some investigations, we found that the discrepancy between the
simulation and theory results was solely attributable to the difference
between the discrete lattice structure of the Fourier space used in
the simulations, and the continuum of Fourier modes used in the
numerical integrals. A detailed discussion of this is presented in
Appendix~\ref{AI1}. In that section we also show that as the
simulation box size is increased, the theory and simulation results
converge. Further, as is shown in Appendix~\ref{AI2} the theory
predictions are sensitive to the lower limit $k_{\rm min}$. In
applying this to the real Universe, we suggest letting $k_{\rm
  min}\rightarrow0$.


\begin{figure}
\centering{
\includegraphics[width=8cm]{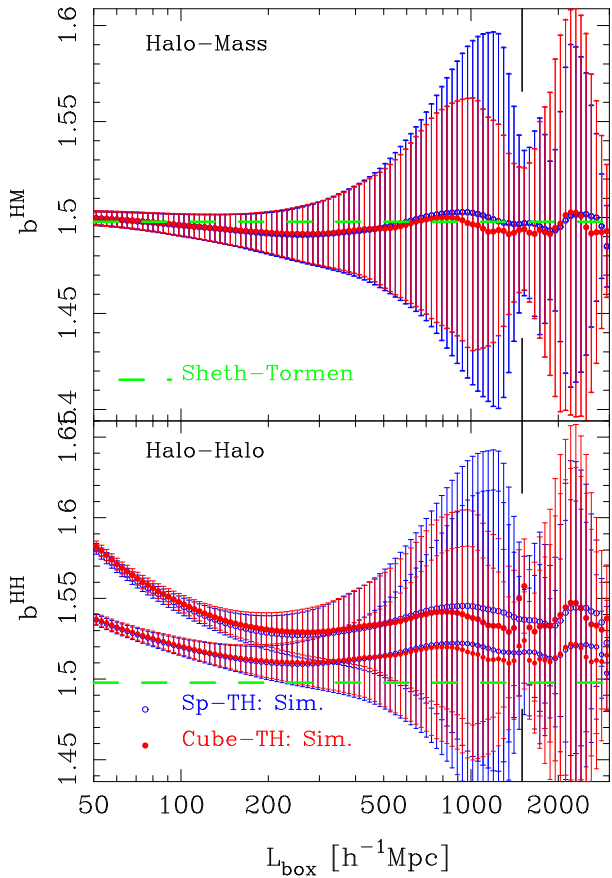}}
\caption{\small{Comparison between the halo bias measured from the
    simulations and the Sheth-Tormen linear theory predictions as a
    function of the sample volume length. The symbols are as in the
    previous figure, and the theoretical prediction is represented by
    the dashed green line. The top panel shows the bias derived from
    the halo-matter variance, while the bottom panel shows the bias
    from the halo-halo variance. The lower panel also shows the
    importance of the shot-noise correction on the $b_{\rm hh}$
    measurements: the upper and lower sets of points denote the
    halo-halo bias before and after the shot-noise correction,
    respectively.}
\label{fig:linear_bias}}
\end{figure}


\subsection{Linearity of the bias}\label{V2}


In linear theory, the relation between the variances plotted in
Figure~\ref{fig:sigma_multiple} is given by:
\be
\sigma^2_{\rm hh}(\Vu)=\overline{b}\, \sigma^2_{\rm hm}(\Vu)
=\overline{b}^2 \,\sigma^2_{\rm mm}(\Vu)\ .
\label{eq:bias_num}
\ee
$\overline{b}$ is the average linear bias from \Eqn{eq:def_bias1},
estimated for a single mass bin containing all halos larger than
$10^{13} \Msol$. For the theoretical bias, we use the Sheth-Tormen
model \cite{ShethTormen1999, ShethTormen2002}, averaged over the same
mass bin. All quantities are at redshift 0. Since the bias is $> 1$,
$\sigma_{\rm hh}(\Vu)$ is slightly larger than $\sigma_{\rm hm}(\Vu)$,
which in turn is also slightly larger than $\sigma_{\rm mm}(\Vu)$. At
this level of detail the differences between the curves appear to be
well related to each other as in \Eqn{eq:bias_num}.

To check this more accurately we next estimate the halo bias in the
simulations and compare it directly with the theoretical predictions.
In direct analogy with the Fourier-space bias estimates in
\citet{Smithetal2007}, we construct the following real-space bias
estimates:
\be
\hat{b}_{\rm hm} \equiv 
\frac{\sigma^2_{\rm hm}}{\sigma^2_{\rm mm}} \ ; \ \ 
\hat{b}_{\rm hh} \equiv 
\sqrt{\frac{\sigma^2_{\rm hh}}{\sigma^2_{\rm mm}}},
\ee
where all quantities in the above depend on $L_{\rm box}$.
Figure~\ref{fig:linear_bias} presents the comparison between the
estimates of the linear bias from the simulations and the values
obtained from the Sheth-Tormen formula. The top and bottom panels show
the results for $b_{\rm hm}$ and $b_{\rm hh}$, respectively. Again the
solid red and empty blue circles denote the results from the cubical
and spherical window functions, respectively.  The Sheth-Tormen theory
is represented by the thick green dashed line.

Considering $b_{\rm hm}$ (top panel), the first thing to remark is
that the bias appears extremely flat over all of the scales probed --
for the mean of the realizations the bias relation is linear to better
than 1\% precision. Secondly, the peak-background split model of Sheth
\& Tormen predicts this value astonishingly well: $b=1.498$.
 
Turning our attention to $b_{\rm hh}$ (lower panel), the raw
simulation measurements (upper set of points) indicate that on scales
$L_{\rm box}\geq 200 \Mpc$, the bias displays a weak scale-dependence
and is roughly $\sim3\%$ higher than the Sheth-Tormen
prediction. However, on smaller scales nonlinear effects are apparent
and the overall amplitude is steadily increasing with decreasing
scale, being $\gtrsim 7\%$ higher than the Sheth-Tormen prediction for
$L_{\rm box}=50\Mpc$. The figure also shows the importance of
correcting $\sigma^2_{\rm hh}$ for shot-noise when making estimates of
the bias. The upper and lower set of points in this panel denote the
uncorrected and corrected estimates, respectively. The shot-noise
correction reduces the discrepancy between the simulations and linear
theory to within $\sim 2\%$ for $L_{\rm box}\geq 200 \Mpc$, however
the nonlinearity on smaller scales remains.

Both cubical and spherical window functions yield very
similar results. In the rest of this work we shall employ the
Sheth-Tormen bias, since on the scales of interest we have shown that
it is at worst $<5\%$ compared to the average bias of the
haloes in our simulations.

Finally, we mention that for the analytical results in the next
sections, we shall use: (i) the volume variance measured from the
matter-matter power spectrum with a cubical window function, and not
the analytical variance, given the discrepancy seen in
Figure~\ref{fig:sigma_multiple}. The cubical window function is a
natural choice, since our simulations also have this geometry; (ii)
the Sheth-Tormen bias; (iii) the Sheth-Tormen mass function.  


\begin{figure*}
\centerline{
\includegraphics[width=6.5cm]{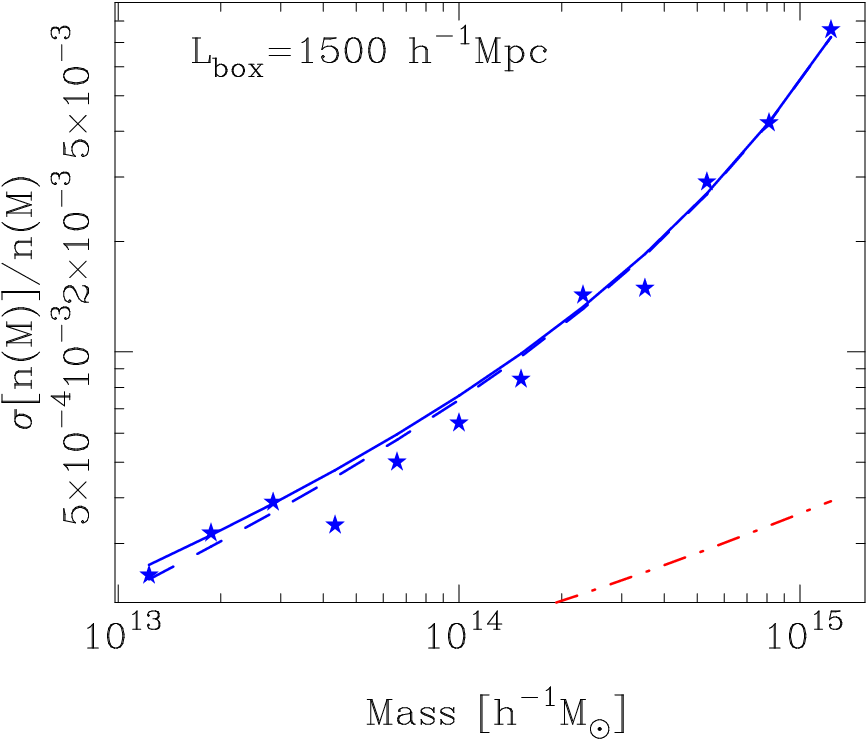}\hspace{0.3cm}
\includegraphics[width=6.5cm]{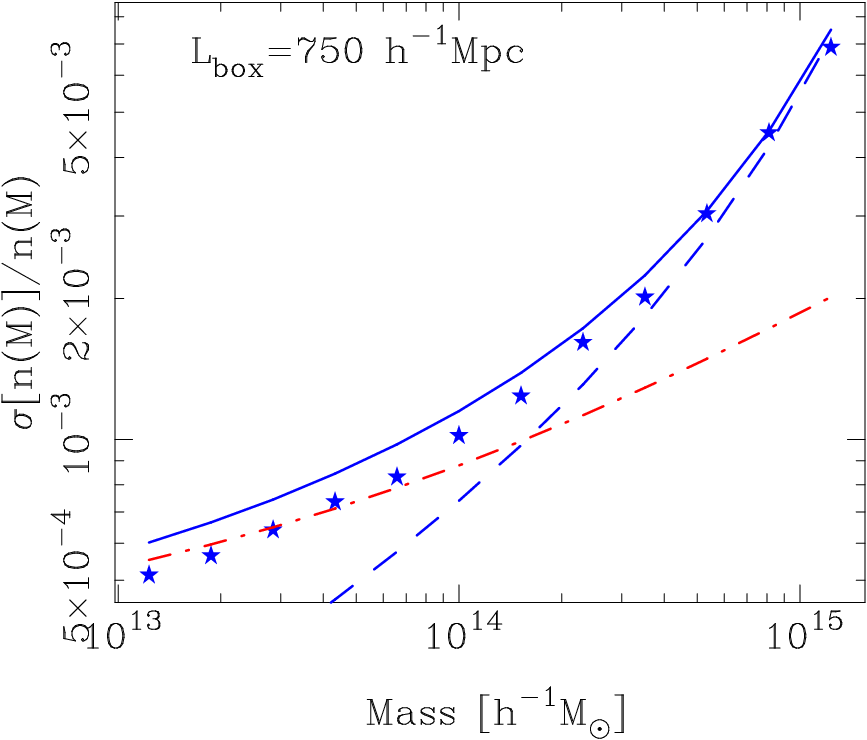}}\vspace{0.3cm}
\centerline{
\includegraphics[width=6.5cm]{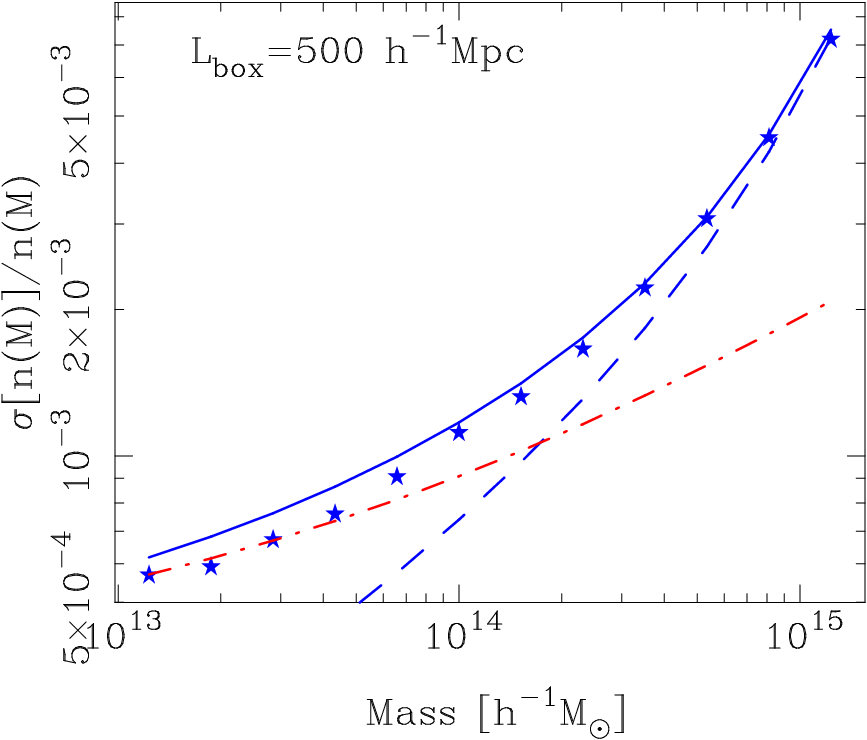}\hspace{0.3cm}
\includegraphics[width=6.5cm]{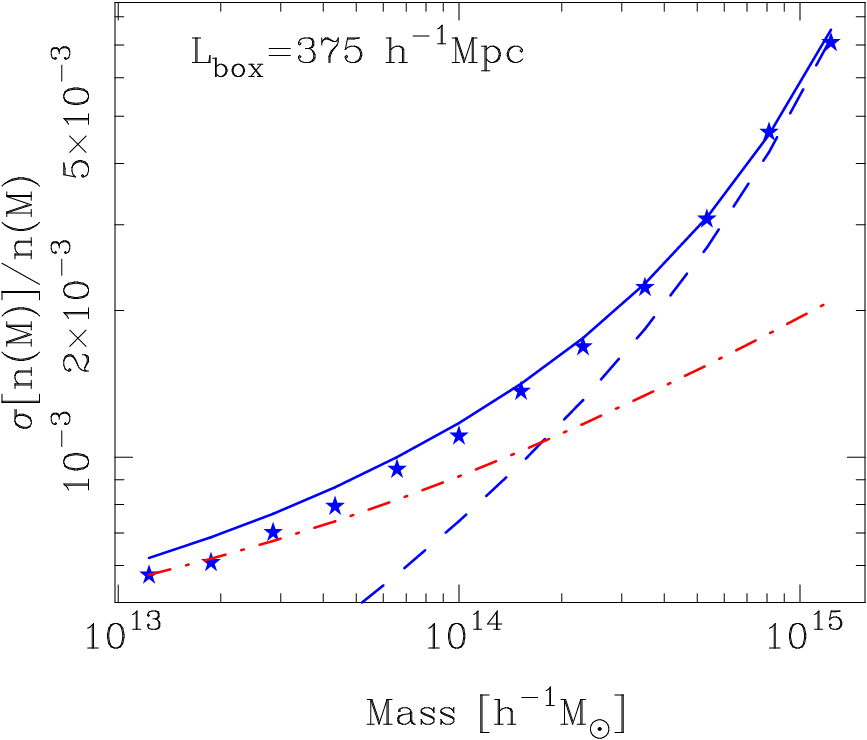}}\vspace{0.3cm}
\caption{\small{{Comparison between the predicted and measured
      fractional error on the halo mass function as a function of halo
      mass. The four panels show the results obtained when the the
      sample volume length is taken to be: \mbox{$L_{\rm
          box}=\{1500\,,750\,,500\,,375\,\}\Mpc$}. In each plot, the
      dashed blue lines denote the fractional Poisson error; the red
      dot-dashed lines denote the pure sample variance error; and the
      solid lines represent the total. All errors have been rescaled
      to a total survey volume of $V=135\Gpccube$.}}}
\label{fig:CompFracErr}
\end{figure*}


\subsection{An estimator for the mass function covariance}\label{V3}


We estimate the mass function covariance matrix from the ensemble of
40 simulations of the fiducial cosmological model, described in
\S\ref{IV}. As we will show shortly, this number of realizations is
insufficient for a reliable estimate of the covariance matrix. In
order to overcome this problem, we have adopted the simple strategy of
subdividing the volume associated with each realization into a set of
smaller cubes. In particular, we divide each dimension of the original
cube by 2, 3, and 4. Hence, each cube of $1500^3\Mpccube$ is
partitioned into 8, 27, and 64 subcubes with corresponding volumes of
$750^3\Mpccube$, $500^3\Mpccube$, and $375^3\Mpccube$,
respectively. The `subcubing' procedure thus provides us with 40, 320,
1080, and 2560 quasi-independent realizations. We note that this
strategy was also adopted by \cite{Crocceetal2010}, who used it to
compute sample-variance error bars on the mass function in the {\tt
  MICE} simulations. However, it has never been employed to compute
the covariance matrix of counts.

One potential disadvantage of this approach, is that the realizations
thus obtained are not perfectly independent, since there will be modes
with wavelength of the order of the initial box size $1500\Mpc$, which
will potentially induce some covariance between the structures in each
set of subcubes. However, as described in Appendix \ref{AII}, we have
checked that this effect is of marginal importance. We shall therefore
treat the measurements in each subcube as providing essentially
independent information. Conversely, the subcubing approach should
actually be thought of as the most relevant scenario, since in the
real Universe there is no cut-off in the power spectrum on scales
larger than the survey.  As we demonstrated in
Figure~\ref{fig:sigma_multiple}, the cut-off scale in the simulations
dramatically affects the behaviour of the density variance on the
scales of the box. Hence, studying the mass function covariance using
simulations that do not account for power on scales larger than the
box modes, may in fact lead to incorrect inferences about the real
Universe.

Our estimator for the covariance matrix can be expressed as
follows. Let $N_{\rm runs}$ be the total number of independent
simulations in the fiducial suite, and $N_{\rm sc}$ the number of
subcubes per simulation that we consider. For each subcube size, we
estimate the average mass function as:
\be \hat{\overline{n}}(M_{\alpha})=\frac{N_{\rm sc}}{V_{\rm sim} \Delta M_{\alpha}}\,
  \frac{1}{N_{\rm tot}} \sum_{i=1}^{N_{\rm tot}} N_{i,\alpha}\,, 
\label{eq:Mave1}
\ee
where we defined $N_{\rm tot}=N_{\rm runs} * N_{\rm sc}$ and $N_{i,
  \alpha}$ is the number of counts in the $i^{\rm th}$ subcube and mass
bin $\alpha$; $V_{\rm sim}=1500^3 \Mpccube$, and $N_{\rm runs}=40$. We
estimate the mass function covariance between mass bins $\alpha$ and
$\beta$:
\ba \hat{{\mathcal M}}_{\alpha \beta} & = &
\left(\frac{N_{\rm sc}}{V_{\rm sim}}\right)^2 \frac{1}{\Delta M_{\alpha} \Delta
  M_{\beta}}\frac{1}{N_{\rm tot}}\sum_{i,j=1}^{N_{\rm tot}}
N_{i,\alpha}N_{j,\beta} \nn \\
& & -\overline{n}(M_{\alpha})\overline{n}(M_{\beta}).\hspace{3.8cm}
\label{eq:Mcov1}
\ea
Note that in the above equation we subtract off the mean mass function
averaged over all subcubes and all realizations in bins $\alpha$ and
$\beta$.  In order to check that the covariance matrix which we
present below, is not affected by our choice of the mean density of
haloes, we recompute it using an alternative method: we determine the
mean density for each realization and subtract it from the counts in
the subcubes of that realization. This alternative is described in
Appendix~\ref{AII}. However, the results obtained from both methods
are consistent.

The covariance matrices of the counts and the mass function are
related through the equation,
\be \hat{C}_{\alpha\beta} =  V^2_{\mu}\Delta M_{\alpha}\Delta M_{\beta}
\hat{{\mathcal M}}_{\alpha \beta} \label{eq:covsim} .\ee

In the following sections we present measurements made at $z=0$. The
mass function analysis is carried out for 12 logarithmically spaced
bins, going from $(10^{13}<M\,[\Msol]<10^{15})$.  Finally, let us make
the clarification that when we refer to `halo mass', we mean the mass
returned from the FoF algorithm.


\begin{figure*}
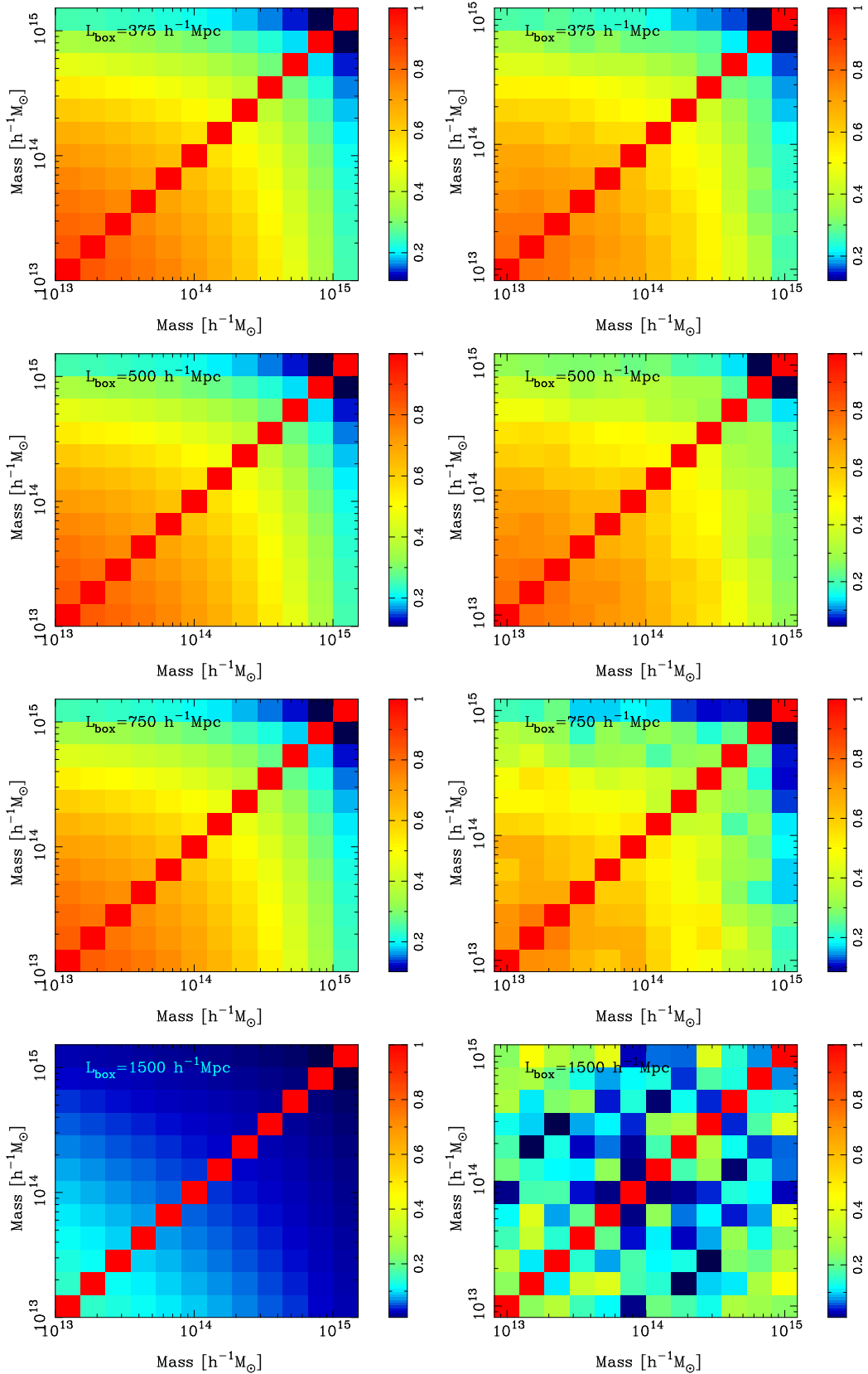

\centerline{
\includegraphics[width=6.7cm]{FIGS/MassFuncCorr_SigVFromSimCube_zHORIZON_Box_375.0_nmbins_12.ps}
\hspace{0.15cm}
\includegraphics[width=6.7cm]{FIGS/SIM_MassFuncCorr_zHORIZON.Box_375.0_nmbins_12.ps}}
\vspace{0.15cm}
\centerline{
\includegraphics[width=6.7cm]{FIGS/MassFuncCorr_SigVFromSimCube_zHORIZON_Box_500.0_nmbins_12.ps}
\hspace{0.15cm}
\includegraphics[width=6.7cm]{FIGS/SIM_MassFuncCorr_zHORIZON.Box_500.0_nmbins_12.ps}}
\vspace{0.15cm}
\centerline{
\includegraphics[width=6.7cm]{FIGS/MassFuncCorr_SigVFromSimCube_zHORIZON_Box_750.0_nmbins_12.ps}
\hspace{0.15cm}
\includegraphics[width=6.7cm]{FIGS/SIM_MassFuncCorr_zHORIZON.Box_750.0_nmbins_12.ps}}
\vspace{0.15cm}
\centerline{
\includegraphics[width=6.7cm]{FIGS/MassFuncCorr_SigVFromSimCube_zHORIZON_Box_1500.0_nmbins_12.ps}
\hspace{0.15cm}
\includegraphics[width=6.7cm]{FIGS/SIM_MassFuncCorr_zHORIZON.Box_1500.0_nmbins_12_v2.ps}}
\vspace{0.15cm}
\caption{\small{The correlation matrix of the cluster mass function
    $r_{\alpha\beta}$, i.e. \Eqn{eq:corrcoef}. The left and right
    columns show the results from theory and simulations,
    respectively.  From top to bottom, the size of the sample volume
    is given by: \mbox{$L_{\rm box}=\{375,\,
      500,\,750,\,1500\}\,\Mpc$}. The theoretical predictions for the
    correlation matrix are generated using the estimate of
    $\sigma_{\rm mm}(\Vu)$ measured directly from the
    simulations. Note that in the bottom right panel we plot
    $|r_{ij}|$, so as to maintain the same heat-bar intensity scale as
    in the other plots.}}
\label{fig:CompTheoSim}
\end{figure*}


\begin{figure*}
\centerline{
  \includegraphics[width=16cm]{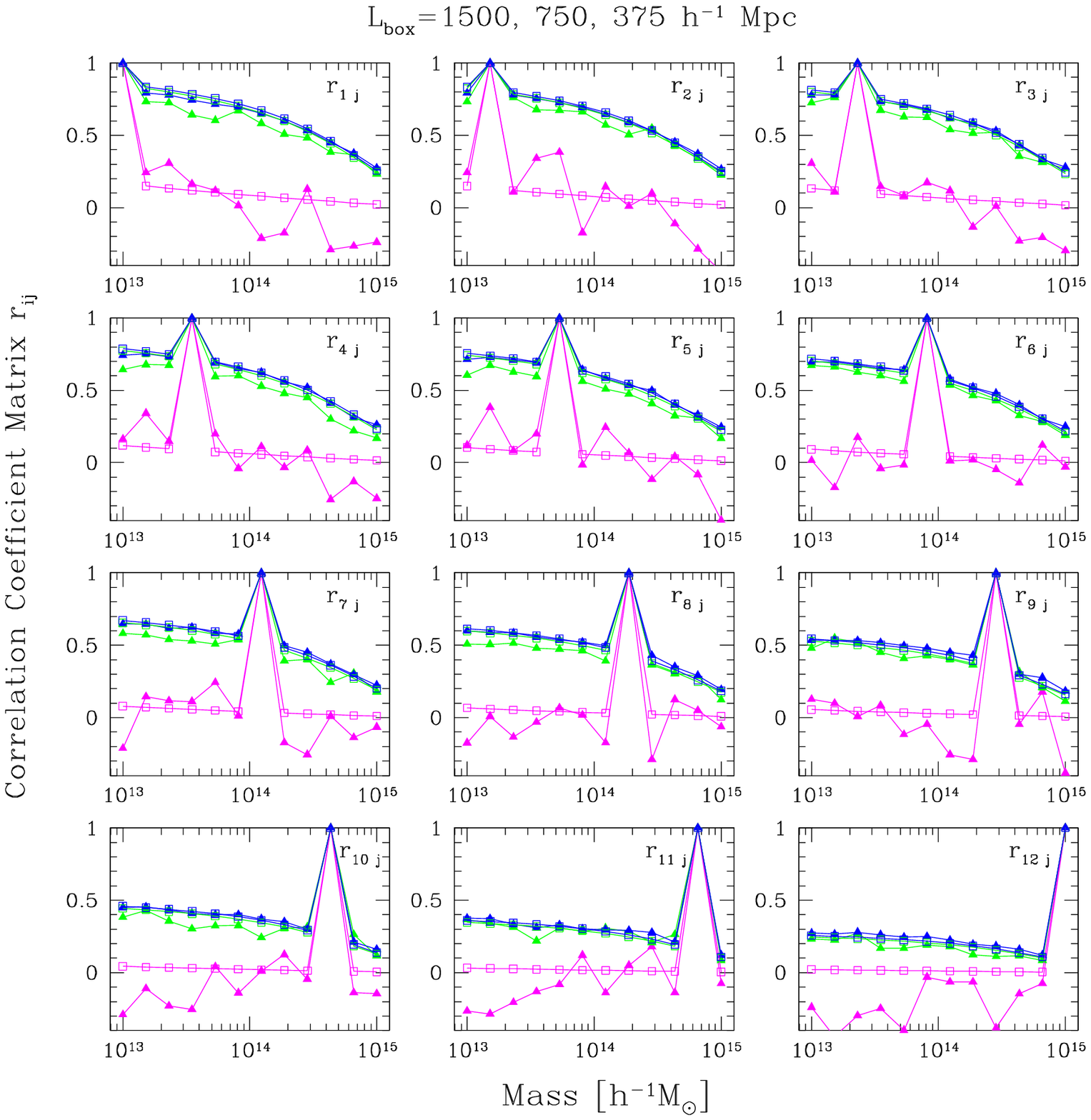}}
\caption{\small{Rows of the cluster mass function correlation matrix
    $r_{\alpha\beta}$ as a function of the mass scale $M_{\beta}$,
    with $M_{\alpha}$ fixed. Each of the 12 panels shows the results
    for one of the 12 rows of $r_{\alpha\beta}$. In all panels, the
    theoretical predictions and the measurements from the simulations
    are denoted by the empty squares and solid triangle symbols,
    respectively.  The magenta, green, and blue colours represent the
    sample volume sizes $L_{\rm box}=\{1500,\,750,\,375\} \Mpc$,
    respectively.  }
\label{fig:CompCorrCoef}}
\end{figure*}


\subsection{Measurements: variance}\label{V4}


Figure~\ref{fig:CompFracErr} presents the fractional errors on the
mass function, $\sigma[n(M)]/n(M)$, from both theory and simulations,
for the subcube sizes mentioned in \S\ref{V3}. In all panels the
points denote the measurements from the simulations. The theoretical
predictions of \Eqn{eq:massfuncov3}, are estimated for a single
realization of given size $L_{\rm box}$, following the recipe at the
end of \S\ref{V2}. Then the variance is rescaled by $1/N_{\rm tot}$,
so that the fractional errors in all four panels correspond to a total
volume of $135\, \Gpccube$.

The agreement between the theory and the measurements is very good,
with a slight difference at the low-mass end for the subcubes
considered. This difference does not occur when the estimate is made
using the full simulation boxes to estimate the variance (see top left
panel of Figure~\ref{fig:CompFracErr}). We also note that the Poisson
model (dashed lines) agrees well with the simulations at the high-mass
end.  However, at lower masses, the variance becomes dominated by the
sample variance, as given by the first term of
\Eqn{eq:massfuncov3}. For a mass bin $\alpha$ the latter is simply:
\be \frac{\sigma[n(M_{\alpha})]}{n(M_{\alpha})}\approx \bar
b_{\alpha}\,\sigma(\Vu)\ ,\ee
and this is denoted in Figure~\ref{fig:CompFracErr} by the dot-dashed
lines.

On comparing all four panels, we observe that with the exception of
the first panel with $L_{\rm box}=1500\Mpc$, the results are almost
indistinguishable. This is quite interesting, since for these subcube
volumes, \Fig{fig:sigma_multiple} shows $\sigma(\Vu)$ to be a
decreasing function of $L_{\rm box}$. For a given mass bin we would
expect the errors for the smaller subcube measurements to be
significantly larger. This is indeed the case, but the fact that we
use the variance on the mean, i.e. we divide by $\sqrt{N_{\rm tot}}$,
leads to results that are very similar.

The slight difference between the measurements and the predictions is
not easy to understand, since for the theoretical estimation we use
$\sigma(\Vu)$ measured from the simulations. This is done for all
subcubes, so we do take into account that modes with wavelength larger
than the subcube size may contribute to the covariance in the
subcubes. The limit is set by the size of the original simulation box
$L_{\rm sim}=1500 \Mpc$. However, the bias is the Sheth-Tormen
prescription, which Figure~\ref{fig:linear_bias} shows to be slightly
lower than the one measured from the halo-halo power spectrum. This
effect might be more pronounced for the small-mass bins, but more work
is needed here to arrive to a definitive conclusion, and we defer this
to a future study.

Before moving on, we note that this startling agreement for the
fractional errors on the mass function was noted before by
\citet{Crocceetal2010}. In that work the variance on a given subcube
scale was computed theoretically using the linear theory variance in a
spherical top-hat taken to have the same volume as the subcube (see
earlier discussion in \S\ref{III}). These authors pointed out that
when using an ensemble of simulations with no subcubing the theory
over-predicted the measurements. Here we have shown that there is no
conflict between the theory and the measurements, if one uses the
volume variance measured from the simulations.


\subsection{Measurements: covariance}\label{V5}


Figure~\ref{fig:CompTheoSim} presents the theoretical mass function
correlation matrix from \Eqn{eq:corrcoef} versus the measured one. The
left panels show the predictions, obtained in the same way as in
Figure~\ref{fig:CompFracErr}, and the right ones the
measurements. From top to bottom, the following subcube sizes are
considered: $L_{\rm box} = 375,\, 500,\, 750,\, 1500\, \Mpc$.

The figure reveals a remarkable agreement between measurements and
theory: the trend observed in Figure~\ref{fig:CompFracErr} is also
present here, with the predictions marginally larger than the
measurements in some of the mass bins. We find that the measurements
are strongly covariant: for clusters with $M\lesssim
<3\times10^{14}\Msol$, the cross-correlation coefficeint is
$r\gtrsim0.5$. Only for the highest-mass clusters does the covariance
matrix become close to diagonal. The exception is for the ensemble of
cubes with $L_{\rm box}=L_{\rm sim}$. In this case the realizations
appear to be only weakly correlated with $r\gtrsim0.1$, for
$M\lesssim3\times 10^{13}\Msol$. However, as was discussed above and
shown in Figure~\ref{fig:sigma_multiple}, the behaviour of
$\sigma(\Vu)$ at the simulation box scale is not representative for
the real Universe, owing to the absence of power on larger scales. Had
we run a simulation of larger volume, then the volume variance on the
scale $L_{\rm box}=1500 \Mpc$ would have been significantly larger.

Figure~\ref{fig:CompCorrCoef} presents the same information as
Figure~\ref{fig:CompTheoSim}, but in a more quantitative format. The
plot has 12 panels, with each panel depicting a single row from the
correlation matrix, i.e. $r_{ij}(M_i,M_j)$ vs $M_{j}$, with $M_{i}$
fixed. In this plot the solid triangles denote the measured
correlation coefficient, while the empty squares represent the theory
predictions. For clarity, we show results only for the box sizes
$1500,\,750,\,375 \Mpc$, represented by the magenta, green, and blue
symbols respectively. It is clear from this figure too that the theory
predictions and the measurements are in remakably good agreement. On
comparing the correlation coefficient for different subcube sizes, we
again note the similarity of these results, despite the variation in
$\sigma(\Vu)$: just as in \Fig{fig:CompFracErr}, the covariance on the
mean leads to the observed similarity. The exception is for the
$L_{\rm box}=L_{\rm sim}$ cubes, and we offer the same explantion for
this as noted above.

We conclude this section by stating that \Eqn{eq:massfuncov3} gives a
very reliable prediction for the mass function covariance, provided
one employs the true variance within the volume.


\section{Cosmological information from the mass function}\label{VI}


In this section we examine how the cosmological information content of
the cluster mass function changes, when we exchange the standard
Poisson assumption for the more complex likelihood models of
\Eqns{eq:likelihood1}{eq:likelihoodG2}.


\subsection{Fisher information}\label{VI1}

In all cases we shall use the standard definition of the Fisher
information \citep[for an excellent review of Fisher matrix techniques
  in cosmology see][]{Heavens2009}:
\be F_{p_ap_b} = - \left< \frac{\partial^2 
\ln {\mathcal L}}{\partial p_a\partial p_b}\right> \ ,\ee
where $p_{a}$ and $p_b$ are elements of the cosmological model
parameter set upon which the likelihood depends. From the Fisher
matrix, one may obtain an estimate of the marginalized errors and
covariances of the parameters:
\be \sigma^2_{p_ap_b} = [F^{-1}]_{p_ap_b},
\label{eq:marg_error}
\ee
as well as the unmarginalized errors: 
\be \sigma_{p_a} = [F_{p_ap_a}]^{-1/2} .
\label{eq:unmarg_error}
\ee
%

\subsection{The Poisson Fisher matrix}\label{VI2}


In the case of Poisson errors for each cell and mass bin, then using
\Eqns{eq:defPoisson}{eq:likelihoodP} we write:
\ba 
\ln {\mathcal L} & = & \sum_{i,\alpha} \ln P(N_{i,\alpha}|\overline{m}_{i,\alpha}) \nn \\
& = & \sum_{i,\alpha} \left[-\overline{m}_{i,\alpha} + 
N_{i,\alpha}\ln  \overline{m}_{i,\alpha} - \ln N_{i,\alpha}! \right] \ .
\ea
On partially differentiating the above expression with respect to
parameters $p_a$ and then $p_b$, and on performing the ensemble
average, one finds:
\be 
F^{\rm Poisson}_{p_ap_b} = \sum_{i,\alpha} \frac{\partial\overline{m}_{i,\alpha}}{\partial p_a}
\frac{\partial\overline{m}_{i,\alpha}}{\partial p_b}\frac{1}{\overline{m}_{i,\alpha}}\ .
\label{eq:FisherPoisson}
\ee
%

\begin{figure*}
\centerline{
\includegraphics[width=15cm]{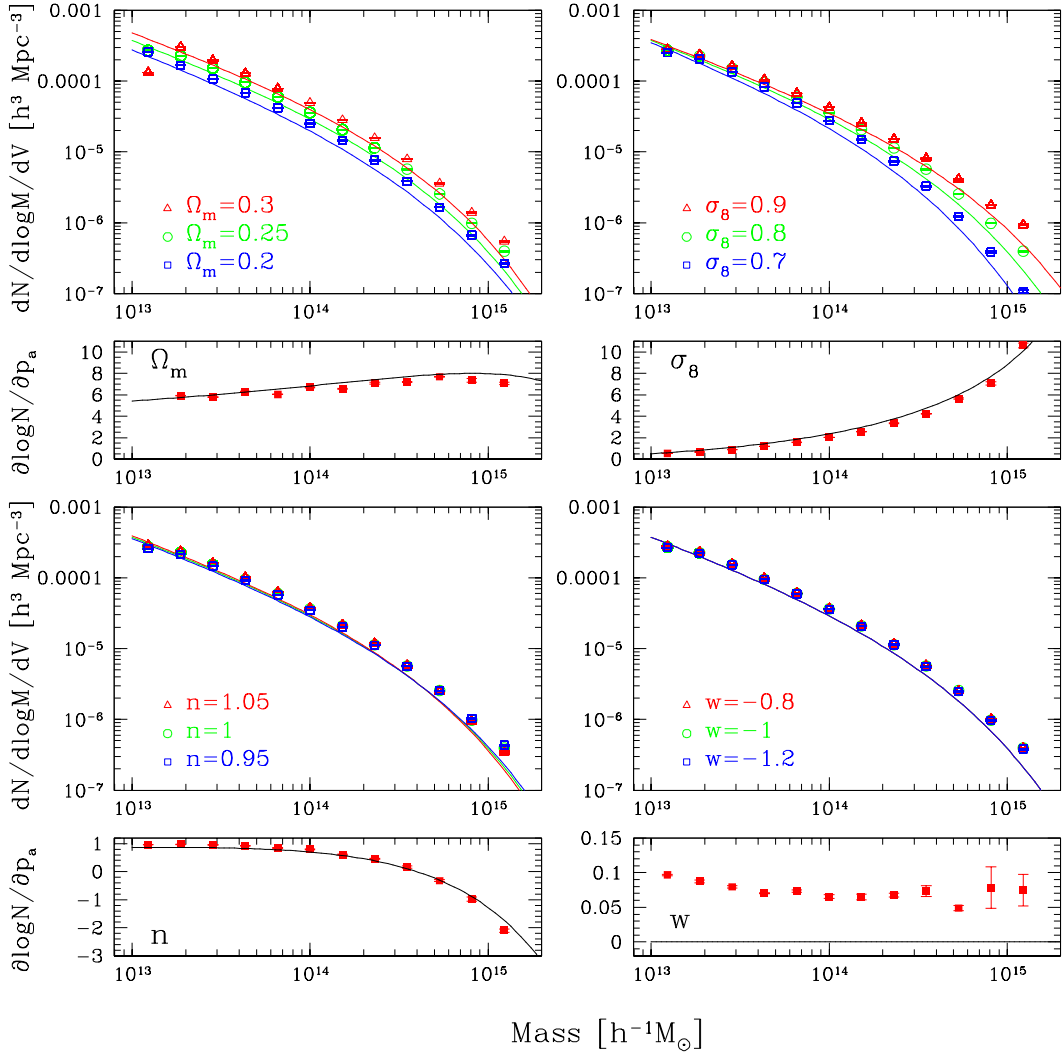}}
\caption{\small{ {\bf Top section of each panel:} Dependence of the
    $z=0$ cluster mass function on cosmology, as a function of cluster
    mass. Symbols denote measurements from the simulations and lines
    depict the \citet{ShethTormen1999} mass function. The green colour
    represents the fiducial model, whereas the red/blue colours are
    for the plus/minus variations in the parameters. {\bf Bottom
      sections:} Logarithmic derivatives of the cluster number counts
    with respect to the considered parameters, as a function of
    cluster mass. Points with errors denote measurements from the
    simulations (c.f.~ Eq.~\ref{eq:logderiv}), the error bars being on
    the mean. Lines denote the Sheth-Tormen predictions.
  }
\label{fig:mf_derivatives}}
\end{figure*}


\subsection{The Gaussian Fisher matrix}\label{VI3}


As was shown earlier, in the case of the full likelihood model for the
counts (c.f. \Eqn{eq:likelihood1}), we expect the Fisher matrix to
be significantly modified from the Poisson case in the region of many
counts per mass bin.  In this limit, the likelihood is given by
\Eqn{eq:likelihoodG2}, and we have the standard result for the Fisher
information for a Gaussian likelihood \citep{Tegmarketal1997}:
\be 
{\mathcal F}_{p_ap_b}^{\rm Gauss}= 
\frac{1}{2} 
{\rm Tr}\left[
{\bf S}^{-1}\frac{\partial{\bf S}}{\partial p_a}
{\bf S}^{-1}\frac{\partial{\bf S}}{\partial p_b}\right]
+\frac{\partial\overline{\bf m}}{\partial p_a}^{T}{\bf S}^{-1}
 \frac{\partial\overline{\bf m}}{\partial p_b}\ .
\ee
%


\subsection{The Gauss-Poisson Fisher matrix}\label{VI4}

LH04 developed an approximation for the Fisher matrix, which
interpolates between the correct forms for the information in the
limit of rare peaks and sample-variance-dominated counts. Their
expression is:
\be 
{\mathcal F}_{p_ap_b}^{\rm G+P} \approx  
\frac{1}{2} 
{\rm Tr}\left[
{\bf C}^{-1}\frac{\partial{\bf S}}{\partial p_a}
{\bf C}^{-1}\frac{\partial{\bf S}}{\partial p_b}\right]
+\frac{\partial\overline{\bf m}}{\partial p_a}^{T}{\bf C}^{-1}
 \frac{\partial\overline{\bf m}}{\partial p_b}
\label{eq:FisherLH}\ ,
\ee
where ${\bf C} = \overline{\bf M}+{\bf S}$, and $ \overline{\bf M}$ is
a diagonal matrix with the elements $\overline{m}_{i, \alpha}$, as
defined in \S\ref{II}.


\begin{figure*}
\centerline{
\includegraphics[width=15cm]{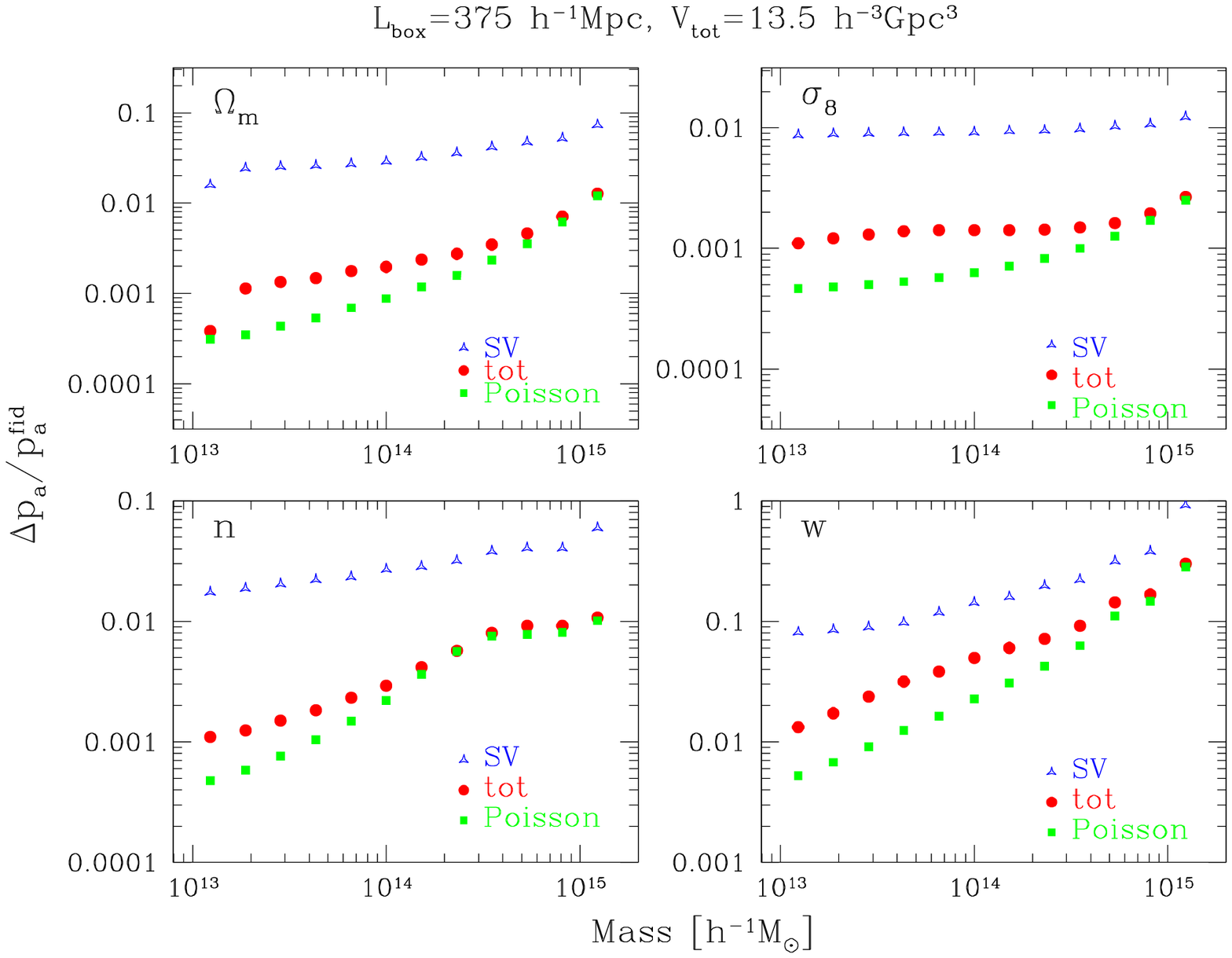}}
\caption{\small{Fractional Fisher-matrix errors on the cosmological
    parameters, as a function of the minimum cluster mass used. The
    four panels show the results for the cosmological parameters:
    $p_{\alpha}\in\{\Omega_m,\,\sigma_8,\,n,\,w\}$. In all panels, the
    symbols show the estimates made from the $N$-body simulations,
    with varying assumptions about the form of the cluster likelihood
    function. Solid green squares denote the Poisson errors obtained
    with \Eqn{eq:FisherPoisson}; solid red circles denote the errors
    obtained from the second term of \Eqn{eq:FisherLH}; blue
    triangular shaped symbols denote the errors derived from the
    trace-term in \Eqn{eq:FisherLH}.  }
\label{fig:errorsU}}
\end{figure*}


\subsection{Estimating Fisher matrices from simulations}\label{VI5}


In order to evaluate all of the expressions for the Fisher matrices
presented in the previous sections, we require knowledge of three
quantities: the partial derivatives of the mean counts with respect to
the parameters $\partial\overline{\bf m}/\partial p_a$; the inverse of
the total covariance matrix ${\bf C}^{-1}$; and the derivative of the
sample variance covariance matrix $\partial{\bf S}/ \partial p_a$. In
this section we shall use numerical simulations to directly evaluate
all of these quantities.

We first measure the halo mass function for each of the variational
cosmologies described in \S\ref{IV}. With this information we are then
able to numerically obtain the derivatives $\partial\overline{\bf
  m}/\partial p_a$ for the simulated parameters $p_{a}\in\{\Omega_m,
\sigma_8, n, w\}$. When computing the mass function derivatives, we
reduce the effects of cosmic variance on the estimates, using the
fact that the first 4 simulations of the fiducial cosmology have
matched initial conditions with the 4 variational-cosmologies
simulations. Hence, our reduced-cosmic-variance estimator for the
derivatives can be written:
\be 
\frac{\partial \bar{n}_\alpha}{\partial p_b}=\frac{\bar{n}_\alpha}{N_{\rm
    var}} \sum_{r=1}^{N_{\rm var}}  \frac{\partial \log \bar{n}^{(r)}_\alpha}{\partial p_b}, 
\label{eq:deriv}
\ee
where $\bar{n}_\alpha$ is the average mass function for mass bin
$\alpha$, estimated from all 40 independent realizations; $N_{\rm
  var}=4$ is the number of the variational simulations; $r$ denotes
the simulation realization going from $1$ to $N_{\rm var}$;
$\bar{n}^{(r)}_\alpha$ is the mass function in the fiducial case,
estimated for each of the 4 realizations that have matched initial
conditions to the variational-cosmologies realizations (for an explict
defintion of this see \Eqn{eq:Mave2}). The logarithmic derivatives are
estimated as:
\be \frac{\partial \log \bar{n}^{(r)}_\alpha}{\partial
  p_b}=\frac{\bar{n}^{(r)}_{\alpha}(p_b+\Delta_{b})-\bar{n}^{(r)}_{\alpha}(p_b-\Delta_b)}{2
  \Delta_b \bar{n}^{(r)}_\alpha(p_b)} \, .
\label{eq:logderiv}
\ee
Note that since we estimate $\partial\overline{\bf m}/\partial p_a$
using double-sided derivatives, we may take larger step sizes in the
parameters to compute the derivatives than would be allowed for single
sided derivatives \citep{Eisensteinetal1999}. For the former case, the
errors in the derivatives are of quadratic order in the step size:
i.e. $\Delta[\partial\overline{\bf m}/\partial p_a]\approx(\Delta
p_a)^2\partial^3\overline{\bf m}/\partial p_a^3/6$. Thus parameter
step sizes of 20\% and 10\% should correspond to relative errors of
roughly 4\% and 1\% in the derivatives, respectively. In actuallity,
the true accuracy of the derivatives also depends on the value of the
third partial derivative.

In Figure~\ref{fig:mf_derivatives} we show simulation measurements of
the average mass functions for the fiducial and variational
cosmologies. This figure makes very clear not only the sensitivity of
the mass function to the cosmological parameters considered, but also
the halo mass range over which most of it occurs. Changes in
$\Omega_m$ and the slope of the primordial power spectrum $n$ impact
the mass function for the whole range of halo masses. The low-mass end
is less sensitive to variations in $\sigma_8$, while the dark energy
equation of state parameter $w$ barely affects the mass function.

In the smaller panels of Figure~\ref{fig:mf_derivatives} we show the
derivatives of the halo abundance, estimated using
\Eqn{eq:logderiv}. The error bars are computed as errors on the mean
of the $N_{\rm var}=4$ realizations, as they are also for the mass
functions in the larger panels. The $\Omega_m$-derivative is almost
constant and large for all bins, while the $\sigma_8$ one
monotonically increases from 0 at the low mass end to a large value at
the high mass end. The behaviour of the spectral-index-derivative is
quite interesting, as it changes sign at $M\sim3\times10^{14}\, \Msol$
and becomes negative at the high mass end. Its overall variation is
not as large as in the case of $\Omega_m$ and $\sigma_8$, which will
be better constrained by the halo abundance. 

Another interesting finding of this exploration concerns the
$w$-derivative, which should be $0$ at redshift $0$ according to
linear theory and the Sheth-Tormen mass function. We find it to be
small and positive, $\sim0.05$, for most of the mass range considered,
and rising slightly to $\sim0.1$ at the low-mass end. The
$w$-derivative does not vanish because in reality the mass function
depends on the full nonlinear growth history. This encompasses the
growth of structure at all redshifts, and thus makes the present day
halo abundance sensitive to $w$. These results are consistent with the
findings in earlier studies
\citep{LinderJenkins2003,Jenningsetal2010}.

We next follow the recipe of \S\ref{V3}, to estimate the covariance
matrices in each of the variational cosmological models. From these
estimates we are then able to form the partial derivatives of the
covariance with respect to the cosmological parameters: $\partial{\bf
  C}/ \partial p_a$. Again, as was done for $\partial\overline{\bf
  m}/\partial p_a$, we take advantage of the matched intial conditions
to reduce the cosmic variance on the estimates of the partial
derivatives of the covariance matrix.


\subsection{Forecasted errors}\label{VI6}

Having obtained all of the necessary ingredients we are now in a
position to evaluate the Fisher information directly from the
simulations.  

Figure~\ref{fig:errorsU} shows the cumulative fractional Fisher
errors, $\Delta p_a/p_a^{\rm fid}$, estimated using
\Eqn{eq:unmarg_error}, as a function of the minimum cluster mass, and
for the four cosmological parameters that we consider. The results
obtained for the various subcube sizes are almost identical with the
exception of the case where $L_{\rm box}=L_{\rm sim}$: as explained
earlier, the underestimate of the variance on scales of the simulation
makes the estimate of the mass function covariance, and hence the
Fisher errors unrealistic. For brevity we shall present only the
findings for $L=375 \Mpc$, which we consider very reliable.

For our fiducial survey, we adopt parameters relevant for future
all-sky $X$-ray cluster surveys, such as eROSITA
\citep{eROSITA2010short}. This mission will be able to target
intermediate mass range clusters, and not just the most massive
objects in the Universe as is the case for current and past surveys.
We adopt a total survey volume of $V\sim13.5 \Gpccube$, and we rescale
our measured covariance matrices to this volume. For this comoving
volume at $z=0$, we find in the simulations approximately
$4.5\times10^{6}$ halos in the mass interval $[1,5]\times10^{13}
\Msol$, $5.4\times10^{5}$ halos in the interval $[0.5, 1]\times
10^{14} \Msol$, $2.3\times10^{5}$ halos in the interval $[1,6.5]
\times 10^{14}\Msol$, and 8000 halos with masses larger than the
latter limit.

In \Fig{fig:errorsU} the solid green squares denote the results
obtained for the Poisson Fisher matrix, as given by
\Eqn{eq:FisherPoisson}.  The solid red circles denote the errors
resulting from the second term of \Eqn{eq:FisherLH}. The blue
triangular-shaped symbols denote the errors obtained from only the
trace-part of \Eqn{eq:FisherLH}, where instead of ${\bf S}$, we have
used the covariance matrix from our simulations $\hat{{\bf C}}$
(c.f. \Eqn{eq:covsim}). We do not expect that replacing $\hat{{\bf
    C}}$ with ${\bf S}$ will change our conclusions concerning the
information carried by this term, except to possibly make the errors
larger.

As expected, for all of the cosmological parameters considered, the
fractional errors obtained from the Poisson approximation are
smallest. Including the full covariance matrix, as in the second term
of \Eqn{eq:FisherLH}, reduces the amount of information, and this
results in a significant increase in the fractional errors. For the
case of $M_{\rm min}\sim 10^{13} \Msol$, the errors are roughly a
factor of $\sim$3 larger when the full-covariance is used as opposed
to the Poisson case. When $M_{\rm min}\sim 10^{14}\Msol$, the errors
are only a factor of $\sim$2 worse. For the rarest objects, where the
covariance becomes almost diagonal, the errors from the two methods
are very similar. We find that the trace part of \Eqn{eq:FisherLH}
contributes negligibly to the information, and if this term is taken
separately, it yields errors that are roughly one order of magnitude
larger than those from the second term. 

Let us explore the consequences of this last result a little
further. Consider the Fisher matrix given by \Eqn{eq:FisherLH}, if the
first term on the right-hand-side is negligable, then the information
about each cosmological parameter enters the system only through the
derivatives of the model mean with respect to the parameters. Since
the model here is the mean counts, the bias provides no
information. However, the amplitude of the elements of the information
matrix can be modulted by the inverse covariance matrix.  Owing to the
fact that increasing the elements of the covariance matrix only leads
to a smaller inverse covariance, we thus conclude that, adding the
variance from the bias can only ever decrease the Fisher
information. However, as discussed in \citet{LimaHu2005}, the
information content of the first term of \Eqn{eq:FisherLH}, becomes of
great importance in the presence of a scatter between the true and
observed mass.

Note also that the cumulative dependence of the errors on the mass
bins, can partly be understood by examining the behaviour of the
derivatives as shown in \Fig{fig:mf_derivatives}. The errors flatten
out at those points in the mass range where the derivatives of the
parameters are close to 0, as in the case of $\sigma_8$ at the
low-mass end, or $n$ at masses $\sim3\times 10^{14}\, \Msol$.

Finally, we emphasize that the forecasts that we make above are to
illustrate the importance of going beyond the Possion likelihood
approximation and should not be taken as serious predictions for a
potential survey. The cosmological dependence that we have considered
here arises strictly from the mass function. In order to make a
realistic forecast we would have to take into account a number of
observational factors: realistic survey geometries; the evolution of
the mass function with redshift; the evolution of the volume element
with the cosmological model; and the evolution in the minimum
detectable mass at each redshift; and a scatter in the relation
between the observed mass proxy and the true cluster mass
\citep[see][for an example of forecasting weak lensing cluster
  counts.]{MarianBernstein2006}.


\section{Summary and conclusions}\label{VII}


In this paper, we have studied the covariance of the halo mass
function, and the cosmological information content of such data. We
adopted a two-line attack on these problems: the first line was
theoretical and we developed an analytic model to explore these
issues; the second was the use of a large ensemble of numerical
simulations to measure directly all quantities of interest.

In \S\ref{II}, we summarized the counts-in-cells formalism
\citep{Peebles1980,HuKravtsov2003}, and developed it for application
to deal with cluster counts in multiple mass bins. We described the
Gauss-Poisson likelihood function for the counts in cells with
multiple mass bins. The expression was analogous to that derived by
\citet{LimaHu2004} for multiple cells and a single mass bin.

In \S\ref{III}, we used this framework to derive a formal expression
for the covariance of the halo mass function and the cross-correlation
coefficient, i.e. \Eqns{eq:massfuncov3}{eq:corrcoef}, respectively.
We found that there were two terms contributing: a Poisson shot-noise
term, which dominates in the limit of rare clusters; and a term
associated with the sample variance, which is dominant for abundant
clusters. This expression is analogous to the results of
\citet{HuKravtsov2003} for multiple cells and a single mass bin. The
expression was found to depend on three quantities: the cluster mass
function; the cluster bias; and the variance in the survey volume.

In \S\ref{IV}, we presented the details of our large ensemble of
numerical simulations: 40 simulations of a fiducial model and 32
simulations of modified cosmological models.

In \S\ref{V}, we made a rigorous comparison of the results from the
theoretical modelling with those obtained directly from the numerical
simulations. We measured the variance of matter and cluster
fluctuations in cells of various sizes and found, for spherical and
cubical top-hat cells, that the simulations and theory predictions
were discrepant for large cell sizes. We showed that this was entirely
attributable to the difference between the discrete lattice structure
of the Fourier space in the simulations, and the continuum of Fourier
modes in the theory integrals. The cubical and spherical top-hat
simulation results were in good agreement, except on the largest
scales where simulation box-scale effects were important. 

We also measured the halo bias in cells of various sizes from the
simulations. We found that the bias from the halo-mass cross-variance
showed very little scale dependence over the range $L_{\rm
  box}=[50,1500]\Mpc$, whereas that from the halo auto-variance showed
significant scale dependence, before and after the shot-noise
correction. We found that the \citet{ShethTormen1999} model was an
excellent fit to the former and a reasonable fit to the latter.

We then measured the covariance of the mass function in the
simulations. To increase the number of realizations, we used the
strategy of subdividing each large simulation volume into a set of
smaller subcubes. We found that the estimated covariances were in
excellent agreement with the theoretical predictions. This was under
the condition that we used the actual variance of mass fluctuations
measured in the simulations.

In \S\ref{VI}, we employed the Fisher matrix formalism to explore the
information content of the cluster counts. Using the more realistic
likelihood functions, we demonstrated numerically that the Poisson
likelihood model only provides a reasonably accurate description of
the data for clusters that are more massive than $M \gtrsim 5.0\times
10^{14}\Msol$. Future surveys that aim to target cluster samples with
masses $M\lesssim 5 \times 10^{14}\, \Msol$, must adopt more
sophisticated likelihood analysis, such as discussed by
\citet{LimaHu2004}, \citet{HuCohn2006} and here in, which take into
account the full covariance matrix of the counts. Otherwise,
significant underestimates of the true errors will occur.

There are a number of possible future directions for the work that we
have presented here. Firstly, as pointed out by \citet{LimaHu2005},
one of the main uses of adopting the counts-in-cells approach is that
it helps to lift the degeneracy between nuisance parameters, which are
involved in calibrating the cluster masses, and the cosmological
parameters. This occurs beacuse the sample varaince depends on the
bias of the clusters, which has a different behaviour with
cosmological parameters than the counts. Whilst we have shown
explicitly that the terms in the Fisher matrix that depend on the
derivatives of the covariance matrix, and hence derivatives of the
bias, do not carry a great deal of cosmological information by
themselves, it will be interesting to see whether for a more realistic
scenario, where one must marginalize over these nuisance parameters,
the self-calibration can be successfully performed to restore the lost
information.

We also note that the counts-in-cells technique has been highlighted
as a powerful means for constraining primordial non-Gaussianity
\citep{Oguri2009,Cunhaetal2010,Marianetal2011}. It is of some
importance to explore this approach using numerical simulations, since
it is not clear whether the extension of the current formalism to such
modified cosmological models works in practice.


\section*{Acknowledgements}
We thank Gary Bernstein and Ravi Sheth for comments on the draft and
Raul Angulo, Martin Crocce, Peter Schneider, Uros Seljak and Yu-Ying
Zhang for useful discussions. We thank V. Springel for making public
{\tt GADGET-2} and for providing his {\tt B-FoF} halo finder, and
R.~Scoccimarro for making public his {\tt 2LPT} code. RES acknowledges
support from a Marie Curie Reintegration Grant, an award for
Experienced Researchers from the Alexander von Humboldt Foundation and
partial support from the Swiss National Foundation under contract
200021-116696/1. LM is supported by the Deutsche
Forschungsgemeinschaft through the grant MA4967/1-1.



\bibliographystyle{apj}


\appendix


\section{The volume variance}
\label{AI}

In this section we investigate the impact of systematic effects on the
volume variance, which arise due to the anisotropic lattice in Fourier
space and also low- and high-$k$ truncation of the matter power
spectrum.


\subsection{Fourier lattice versus a continuum of modes}
\label{AI1}

As was described in \S\ref{III4} the matter variance in the volume is
a key quantity for correctly evaluating the covariance matrix of the
cluster counts. Also, as was shown in \S\ref{V1}, there is a
discrepancy between the theoretical predictions and meaurements in
simulations obtained for $\sigma(\Vu)$. We now investigate the origin
of these discrepancys.

\begin{figure}
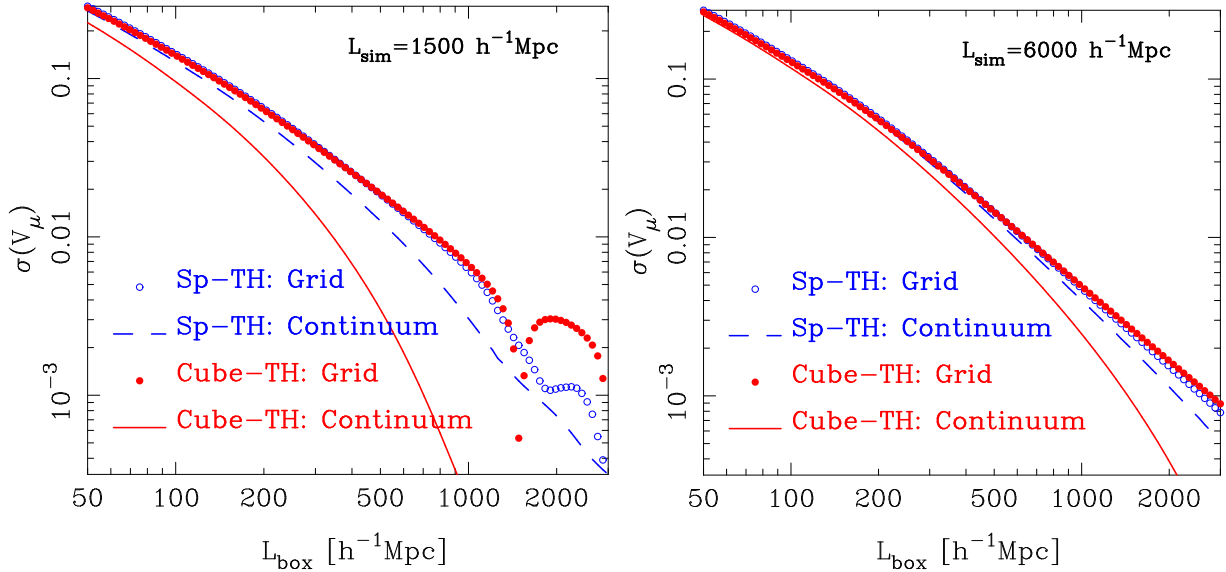

\centering{
  \includegraphics[width=8cm]{FIGS/SigVvsV.TEST.LSim_1500.ps}}\vspace{0.3cm}
\centering{
  \includegraphics[width=8cm]{FIGS/SigVvsV.TEST.LSim_6000.ps}}
\caption{\small{The r.m.s. density variance $\sigma_{\rm mm}(\Vu)$ as
    a function of the sample volume size $L_{\rm box}$.  In each
    panel, blue empty and solid red circles denote theoretical
    predictions made on the Fourier space lattice using the spherical
    and cubical top-hat filter functions, respectively.  The
    predictions made using a continuum of Fourier space modes are
    denoted by the dashed blue and solid red lines, respectively. The
    top panel compares the results when the simulation box size is
    taken to be $L_{\rm sim}=1500 \Mpc$. The bottome panel shows the
    same but for the case where $L_{\rm sim}=6000 \Mpc$.}}
\label{fig:SigVtest}
\end{figure}

We start by examining the importance of the discrete cubical Fourier
lattice, which is used in the estimates from the simulations, and the
continuum of Fourier modes, which is used to evaluate the theory. We
start by generating a Fourier lattice as in the simulations, where
 each lattice point is spaced from the next one, along each dimension,
by $k_{\rm f}=2\pi/L_{\rm sim}$. Then at each lattice point, we
compute the magnitude of the $k$-vector and evaluate the linear theory
power spectrum at that point. $\sigma(\Vu)$ is then obtained as
described in \Eqn{eq:sigmalattice}, by summing up the grid of power
spectra values multiplied by the square of the appropriate window
function.  The top panel of Figure~\ref{fig:SigVtest} shows the
results of this exercise for both the spherical and cubical window
functions. We also compare this to the results obtained from the
theory, assuming a continuum of modes. The results that we find for
the theory evaluated on the cubical mesh, are in remarkably good
agreement with the measurements from the simulations presented in
\Fig{fig:sigma_multiple}.

To be sure that the discrepancy is due to the lattice, we should
expect that as the simulation box size becomes significantly larger,
the results for the lattice should approach those of the continuum.
We test this by regenerating the Fourier lattice, but this time taking
$L_{\rm sim}=6000\Mpc$, and keeping the maximum Fourier mode the same
as before. The results of this exercise are shown in the bottom panel
of \Fig{fig:SigVtest}. We clearly see that the results are now in much
better agreement and for the same cell sizes as in the upper panel.

Thus we are led to conclude that in matching the results from the
simulations we must be mindful to take into account the anisotropic
lattice structure of the Fourier space to obtain accurate comparisons
between the theory and the simulations. This then further justifies
our choice of using the $\sigma(\Vu)$ measured in the simulations to
make the predictions for the covariance of the counts.

Finally, these results also act as a cauationary tale: when
interpreting the results of numerical simulations on very large
scales, we should take more care in asigning the power to the lattice
cells in the initial conditions. We should use methods that supress
this discretization. For instance, it would seem more sensible to
compute the power averaged over a lattice cell and not simply the
power at the lattice cell point. Also including the missing zero modes
may be a more realistic stratergy \citep[see for example][]{Sirko2005}.
 

\subsection{Cut-off scales}
\label{AI2}

In Figure~\ref{fig:sigma_multiple} we evaluated the integrals in
\Eqns{eq:sigmacube}{eq:SigmaV0STH}, keeping the upper and lower bounds
fixed at the values $k_{\rm min}=2\pi/L_{\rm sim}=0.004\kMpc$ and
$k_{\rm max}=2\pi/l_{\rm soft}=100\kMpc$. This was done for a fair
comparison with our simulations, which do not have modes larger than
the simulation box $L_{\rm sim}=1500\Mpc$, nor structures smaller than
the softening scale, $l_{\rm soft}=0.06\Mpc$. We now present a short
discussion of how the mass-fluctuations-variance $\sigma(\Vu)$ depends
on the cell volume and the cut-off scales in the power spectrum.

For the large cell sizes that we are interested in, i.e. $L_{\rm
  box}>50\Mpc$, we find no dependence of $\sigma(\Vu)$ on $k_{\rm
  max}$, for the range of values $k_{\rm max}=[1,100]\kMpc$.

\begin{figure}
\centering{
  \includegraphics[width=10cm]{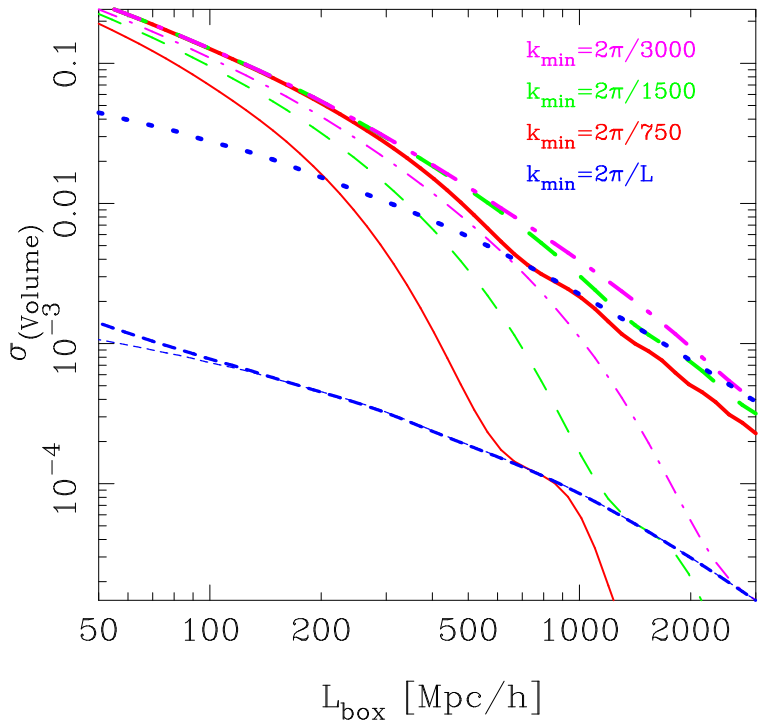}}
\caption{\small{Dependence of the r.m.s. density variance on the
    lower limit $k_{\rm min}$ of the $k$-space integrals, as a
    function of the sample volume size $L_{\rm box}$.  The thin solid
    red, dashed green and dot-dashed magenta lines denote results
    obtained for a cubical top-hat window function, where $k_{\rm
      min}=2\pi/L_{\rm sim}$ with $L_{\rm sim}=\{750,\,1500,\,3000\}
    \Mpc$, respectively. The thick solid, dashed and dot-dashed lines
    represent the same, but for the case where the filter function is
    a spherical top-hat. The thick and thin blue dotted lines denote
    the same as above except this time the lower limit of the
    $k$-space integrals is given by $k_{\rm min}=2\pi/L_{\rm box}$.}}
\label{fig:SigV0}
\end{figure}

For the lower cut-off scale $k_{\rm min}$, the situation appears to be
more complex. In Figure~\ref{fig:SigV0}, we show the result of
computing the mass-fluctuations-variance averaged in cubical and
spherical top-hat volumes, as a function of the cubical cell volume
(recall that we take the radius of the spherical top-hat cell to be
$R=(3/4\pi)^{1/3}L_{\rm box}$).  In the plot we consider the values of
$\sigma(\Vu)$ for three different simulation sizes: $L_{\rm
  sim}=\{750,\,1500,\,3000 \Mpc\}$. These box sizes correspond to the:
solid red, long-dashed green, and dot-dashed magenta lines,
respectively. The thicker/thinner lines in the plot depict the
spherical/cubical top-hat smoothing.


For the case of the spherical top-hat filter, we find that the
variance for $k_{\rm min}=2\pi/750 \kMpc=0.008\kMpc$, is roughly a
factor of $\sim$2 times smaller than the variance obtained when
$k_{\rm min}=2\pi/3000 =0.002\kMpc$.  However, for the case of the
cubical top-hat window function, we find that the difference in the
variance for these same two values of $k_{\rm min}$, is more than an
order of magnitude.

In Figure~\ref{fig:SigV0}, the thick dotted blue curve presents
predictions for $\sigma(\Vu)$ with the spherical window function, but
where we now take the lower limit $k_{\rm min}=2\pi/L$. The thick and
thin dashed blue lines show the same, but for the case of the cubical
filter function. For this case, the thin line is obtained when the
linear theory matter power spectrum is used, and the thicker line
shows the results obtained when the nonlinear power spectrum from {\tt
  halofit} \citep{Smithetal2003} is employed. The differences are very
small. Thus using the linear theory power spectrum for the mass
variance is quite reasonable on these scales. The main point of this
last example, is to show that for large cell sizes, the results for
$\sigma(\Vu)$ are very sensitive to the presence/absence of power on
very large scales.


\section{Convergence of the covariance matrix}\label{AII}


\subsection{Covariances from individual simulations}\label{AII1}


Here we consider an alternate approach to estimating the covariance of
the cluster counts. We are concerned that, if there is a significant
variance of the cluster counts on the scales of the simulation cube,
then by computing the covariance around the mean cluster mass function
averaged over all simulations, we are overestimating the
covariance. To anwer this question, we adopt the stratergy of using
the sub-cubes in a single simulation to make an estimate of the
covariance, and finally we then average these estimates over all the
simulations.

For each simulation run we therefore have:
\be \overline{n}_r(M_{\alpha})=\frac{N_{\rm sc}}{V_{\rm sim} \Delta
  M_{\alpha}}\,\frac{1}{N_{\rm sc}}\sum_{i=1}^{N_{\rm sc}}
N_{i,\alpha}^{(r)}\, ,
\label{eq:Mave2}
\ee
where $N_{i,\alpha}^{(r)}$ is the number of counts in the $r^{\rm th}$
run, $i^{\rm th}$ subcube and $\alpha^{\rm th}$ mass bin.  The
covariance for each run is:
\ba {\mathcal M}^{(r)}_{\alpha \beta} & = & \left(\frac{N_{\rm
    sc}}{V_{\rm sim}}\right)^2 \frac{1}{\Delta M_{\alpha} \Delta
  M_{\beta}}\frac{1}{N_{\rm sc}}\sum_{i,j=1}^{N_{\rm sc}}
N_{i,\alpha}^{(r)} N_{j,\beta}^{(r)} \nn \\
& &  - \overline{n}_r(M_{\alpha}) \overline{n}_r(M_{\beta})\,, \hspace{3.4cm}
\label{eq:Mcov2}
\ea
and the average covariance:
\be
{\mathcal M}_{\alpha \beta} = \frac{1}{N_{\rm runs}} 
\sum_{r=1}^{N_{\rm runs}} {\mathcal M}^{(r)}_{\alpha \beta} \ .
\ee
We have checked that using \Eqns{eq:Mave2}{eq:Mcov2} does not change
the measured mass function covariance in any significant way. We
therefore conclude that the method of estimating the covariance
described in \S\ref{V3}, is not biased by the estimates of the mean
density.


\subsection{The chessboard test}\label{AII2}

When dividing a big simulation box into smaller subcubes, the largest
wavelength modes may affect the observables measured in the smaller
cubes. In the case of clusters, some of the subcubes may have very
different mean counts than others, and in general, the smaller the
sub-boxes, the larger the expected covariance between them. This is
also true for real surveys, which measure observables in a finite
volume of the Universe: some of these observables are impacted by
modes larger than the size of the survey.

In order to check the validity of our approach, we measure the
covariance of the mass function using subcubes that are not adjacent,
and should therefore be less covariant. We shall refer to this as the
`chessboard test', as its 2D analogue would be similar to using only
the white or the black squares of a chessboard to compute the mass
function covariance. This test has the limitation that large mode
correlations can span more than just 2 subcubes, particularly if the
latter are small. Nevertheless, if we find the covariance measured
from the `white' subcubes different from that obtained from the
`black' ones and also different from the all-subcubes-covariance, then
our box-division method is flawed.

\begin{figure}
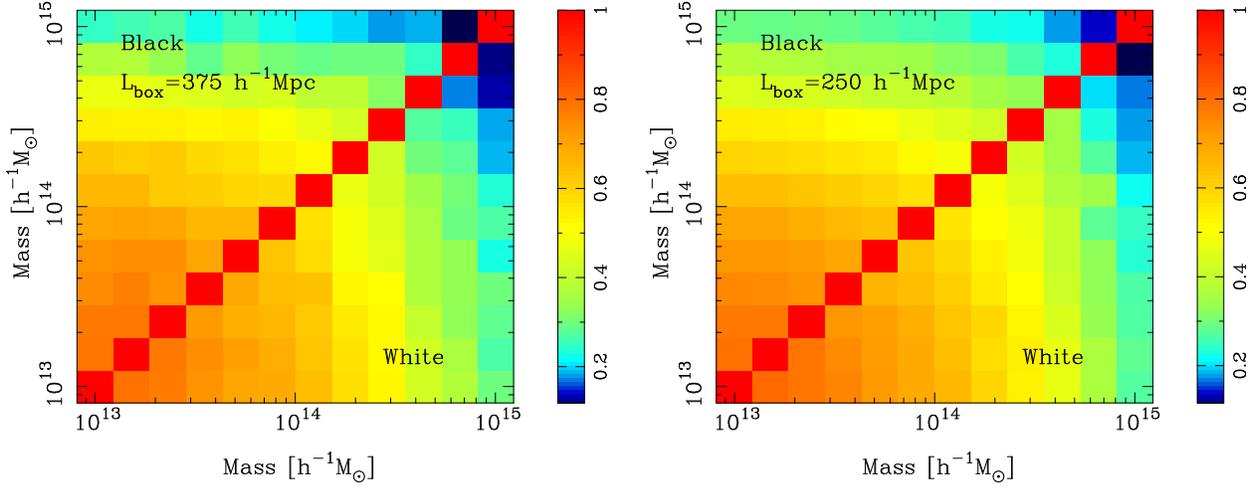

\centerline{
  \includegraphics[width=8cm]{FIGS/CHESS_SIM_MassFuncCorr_zHORIZON.Box_375.0_nmbins_12.ps}
\hspace{0.3cm}
  \includegraphics[width=8cm]{FIGS/CHESS_SIM_MassFuncCorr_zHORIZON.Box_250.0_nmbins_12.ps}}
\caption{\small{The chessboard test: we compare the mass function
    correlation matrix measured from `white' and `black' subcubes--see
    the Appendix text. The results are very similar for both subcube
    sizes considered, $375^3\,\Mpccube$ and $250^3\,\Mpccube$.}
\label{fig:chessboard}}
\end{figure}

We perform this test for the conservative values $n=4,\,6$, i.e. we
consider $4^3$ and $6^3$ subcubes, with volumes $375^3\,\Mpccube$ and
$250^3\,\Mpccube$, respectively. The result is shown in
Figure~\ref{fig:chessboard}. There is no major difference between the
`white', `black', and total mass function correlation matrix
(c.f. \Fig{fig:CompTheoSim}). We conclude that modes with wavelengths
smaller than the size of the subcubes considered here, are not
explicitly responsible for generating the mass function covariance.


\end{document}